\documentclass[prd,nofootinbib,eqsecnum,final]{revtex4}
\usepackage{graphicx,color}
  \usepackage{bm}
   \usepackage{amsmath}
    \usepackage{amssymb}
     \usepackage{pifont}
      \usepackage{simplewick}
\usepackage{tikz}
\usepackage[most]{tcolorbox}
\usepackage{rotating}

\newcommand{\nn}{\nonumber}

\newcommand{\Tr}{\mathrm{Tr}}

\newcommand{\ot}{\leftarrow}

\renewcommand{\(}{\left(}
\renewcommand{\)}{\right)}
\renewcommand{\[}{\left[}
\renewcommand{\]}{\right]}

\renewcommand{\vec}[1]{\bm{#1}}
\newcommand{\fnot}[1]{\not{\! #1}}

\begin{document}
\title{Matching of transverse momentum dependent distributions at twist-3}

\author{Ignazio Scimemi}
\affiliation{Departamento de F\' isica Te\'orica, \\ Universidad Complutense de Madrid,
Ciudad Universitaria, \\ 28040 Madrid, Spain}
\email{ignazios@fis.ucm.es}
\author{Alexey Vladimirov}
\affiliation{Institut f\"ur Theoretische Physik, \\ Universit\"at Regensburg,\\
D-93040 Regensburg, Germany}
\email{alexey.vladimirov@physik.uni-regensburg.de}

\begin{abstract}
We derive the leading order matching of the quark generated polarized transverse momentum dependent (TMD) distributions onto the collinear functions at small values of the transverse distance. Starting from the very definition of the TMD operator and performing the light-cone operator product expansion up to twist-3 order, we evaluate each distribution directly in position space. We primarily consider the cases of Sivers, Boer-Mulders and worm-gear functions. The effects of the TMD process dependence on the matching are explicitly shown. We also discuss the moments of TMD distributions which can be relevant for lattice calculations.
\end{abstract}
\maketitle

\section{Introduction}
\label{sec:introduction}

One of the modern challenges of QCD is represented by the study of the effects of polarization in differential cross sections for semi-inclusive deep inelastic scattering  (SIDIS) and Drell-Yan/vector/scalar-boson production. The cross sections can be factorized in terms of transverse momentum dependent (TMD) distributions~\cite{Collins:2011zzd,Echevarria:2012js,Echevarria:2014rua,Vladimirov:2017ksc} which describe the (transverse and collinear) momentum distribution of quarks and gluons inside a nucleon. The perturbative inputs of the factorization formula play an important role, and presumably should be included at the maximum allowed order to provide the best agreement/prediction of the theory with experiment. The importance of perturbative input
 both at high and low energy, was demonstrated for instance in \cite{Scimemi:2017etj} within the analysis of unpolarized TMD parton distribution functions (TMDPDFs).

Some perturbative parts of the TMD factorization are universal and independent of polarization. It concerns primary the hard coefficient function and evolution kernels, which nowadays are known up three-loop order \cite{Echevarria:2015byo,Li:2016ctv,Vladimirov:2016dll}. The additional parts that require the perturbative input are the actual models for TMD distributions.
 The perturbative computation gives us some relevant information in the limit of large transverse momentum (that is, in the limit of small transverse distances or small-$b$).
  In this limit, the TMD distributions match collinear distributions providing a starting point for phenomenology and   greatly increasing the agreement with high-energy data. Nowadays, the matching for the most part TMD distributions are known, although the information is often difficult to extract from the literature. The matching of TMD distributions to the twist-2 functions is known uniformly at the one-loop level \cite{Gutierrez-Reyes:2017glx}, and some of them are known at two-loop level \cite{Echevarria:2015usa,Echevarria:2016scs,Gutierrez-Reyes:2018iod}. The matching onto the twist-3 functions is less elaborated. An early study was provided in \cite{Bacchetta:2008xw} prior to the  modern formulation of factorization theorem. The main aim of this paper is to uniformly evaluate the matching of quark TMDPDFs at the twist-3 level, using solely the definition of the TMD distribution as it is provided by the TMD factorization theorem. 

The desired matching expressions are given by the first terms of the operator product expansion (OPE) for the TMD operator at small-$b$, or in the vicinity of the light-cone. It can be systematically done starting from the field-theoretical definition of the TMD operator, using the algebra of fields and QCD equations of motion. We recall that the OPE is naturally formulated in the position space, and can be performed without any explicit reference to a particular process. So, the method is universal and allows one to calculate any TMD distribution at any order. Here we consider only the contributions of twist-2 and twist-3 functions, and it covers almost all (7 out of 8) quark TMDPDFs. The gluon distributions, as well as, TMD fragmentation functions (TMDFFs) can be considered in principle in the same fashion. It is important to point out that the OPE does not depend on hadronic states, therefore, many results of this work can be  applied directly, or with a minimal effort, to closely related areas, such as studies of generalized transverse momentum distributions (GTMDs) and Wigner function \cite{Meissner:2009ww,Lorce:2011dv}. Additionally, we think that OPE approach is technically simpler and more systematic in comparison to a direct evaluation and factorization of cross-sections, which is often used see e.g. \cite{Ji:2006ub,Kang:2011mr,Gehrmann:2014yya}.
 
In order to realize our computations we start observing that the operators that define TMD distributions have a  peculiar structure, which distinguishes them from more traditional parton distribution functions (PDFs) and fragmentation functions (FFs), distributions amplitudes and others. Namely, they include half-infinite light-like Wilson lines, and they are geometrically non-compact. Moreover, the direction of the Wilson lines depends on the underline process. This direction is future pointing for production mechanisms (such as fragmentation in semi-inclusive deep-inelastic scattering (SIDIS)), and past pointing for collision mechanisms (such as Drell-Yan process). It gives a superficial process dependence of the TMD distributions, in the form of sign-flip for P-odd distributions \cite{Collins:2002kn}. At the same time,  the collinear distributions are perfectly independent on the process, therefore, any process dependence of TMD operator must reveal itself within OPE. We indeed observe this effect and demonstrate that it appears in the contributions specific for the TMD operators. Exactly these contributions give rise to the famous Efremov-Teryaev-Qiu-Sterman (ETQS) function  \cite{Efremov:1983eb,Efremov:1984ip,Qiu:1991pp,Qiu:1991wg}. Moreover, we have observed that these process-dependent terms of OPE, also contribute to the P-even, and hence process-independent, distributions. Altogether, to our best knowledge, the analysis presented here is the first study of OPE for TMD operators beyond the twist-2 accuracy. As a final result, we obtain all leading order matching expressions for the TMDPDFs in the quark sector.

The definition framework and all relevant TMD distributions and operators are given in sec.~\ref{sec:TMDdefs}. The light-cone OPE for TMD operators up to linear in transverse positions terms is given in sec.~\ref{sec:TMDexpansion}. The parameterization of relevant collinear distributions is presented in sec.~\ref{sec:collinear}. The assembling of OPE and its application to particular distributions is presented in sec.~\ref{sec:matching}. We also present by-product result for Mellin moments of worm-gear function in sec.~\ref{sec:MMworm}. The final results are shown and commented in sec.~\ref{sec:results}.

\section{Definition of TMD distributions}
\label{sec:TMDdefs}
We outline our work for the transverse momentum dependent parton distribution functions (TMDPDFs)
which, in the quark case, 
are defined by the  matrix element~\cite{Tangerman:1994eh,Collins:2011zzd,GarciaEchevarria:2011rb}
\begin{eqnarray}\label{def:TMDPDF_Qop}
\Phi_{q\ot h,ij}(x,\vec b)&=&\int\frac{dz}{2\pi}e^{-ixz(pn)}\langle P,S|\bar T\{\bar q_j\(zn+\vec b\)[\lambda n+\vec b,\pm\infty n]\}T\{
[\pm\infty n,0] q_i(0)\}|P,S\rangle,
\end{eqnarray}
where $\pm \infty n$ indicates different light-cone infinities. The TMD distributions which appear in SIDIS have Wilson lines pointing to $+\infty n$, while in Drell-Yan they point to $-\infty n$. The Wilson lines within the TMD operator are along a light-like direction $n$. Another light-like vector is associated with the large-component of the hadron momentum $P$,
\begin{eqnarray}\label{def:pmu}
p^\mu=(np)\bar n^\mu=P^\mu-\frac{n^\mu}{2}\frac{M^2}{(nP)},
\end{eqnarray}
where $(nP)=(np)$, and $M$ is the mass of hadron ($P^2=M^2$). Together vectors $n$ and $\bar n$ define the scattering plane. The relative normalization of vectors is
\begin{eqnarray}\label{def:n}
n^2=\bar n^2=0,\qquad (n\bar n)=1.
\end{eqnarray}
Thus, any four-vector can be decomposed into the components
\begin{eqnarray}
v^\mu=v^+ \bar n^\mu+v^- n^\mu+v_T^\mu,
\end{eqnarray}
where $v^+=(nv)$, $v^-=(\bar n v)$, and $v_T$ is the transverse component orthogonal to the scattering plane $(v_Tn)=(v_T\bar n)=0$. To specify the reference frame we state that $v^\pm=(v^0\pm v^3)/\sqrt{2}$. 

The transverse components play a special role in our consideration. The transverse subspace is projected out by the transverse part of the metric tensor
\begin{eqnarray}
g_T^{\mu\nu}=g^{\mu\nu}-\frac{n^\mu p^\nu+p^\mu n^\nu}{(np)}.
\end{eqnarray}
There are only two non-zero components, $g^{11}_T=g_T^{22}=-1$. In the following, we also need the transverse part of the Levi-Civita tensor
\begin{eqnarray}
\epsilon_T^{\mu\nu}=\frac{n_\alpha p_\beta}{(np)}\epsilon^{\alpha\beta\mu\nu},
\end{eqnarray}
where $\epsilon^{\mu\nu\rho\sigma}$ is defined in the Bjorken convention ($\epsilon_{0123}=-\epsilon^{0123}=1$). Consequently, we have $\epsilon_T^{12}=-\epsilon_T^{21}=1$, which coincides with the definition \cite{Tangerman:1994eh,Mulders:1995dh,Bacchetta:2006tn}, despite the opposite normalization of the four-dimension $\epsilon$-tensor. The tensor $\epsilon_T^{\mu\nu}$ does not change sign when both indices are down, $\epsilon_{T\mu\nu}=\epsilon_T^{\mu\nu}$, and $\epsilon^{\mu\nu}_T\epsilon_{T,\mu\rho}=\delta^{~~\nu}_{T,~\rho}$. Since the transverse subspace is Euclidian, the scalar product transverse vectors is negative, $v_T^2<0$. In the following, we use the bold font notation to designate the Euclidian scalar product of transverse vectors, i.e. $\vec b^2=-b^2>0$, when it is convenient.

The spin of the hadron is parameterized by the spin-vector $S$,
\begin{eqnarray}
S^2=-1,\qquad (PS)=0.
\end{eqnarray}
The light-cone decomposition of the spin vector is
\begin{eqnarray}\label{def:Smu}
S^\mu=\frac{\lambda}{M}p^\mu-\frac{\lambda}{2}\frac{M}{(np)}n^\mu+s_T^\mu,
\end{eqnarray}
where the helicity $\lambda$ of the hadron  is
\begin{eqnarray}
\frac{(nS)}{(np)}=\frac{\lambda}{M}.
\end{eqnarray}
The vector $s_T^\mu$ is the transverse component of the spin, $s_T^2=\lambda^2-1$. With the help of $\epsilon_T$-tensor we can introduce another useful (axial) vector
\begin{eqnarray}
\tilde s^\mu_T=\epsilon^{\mu\nu}_TS_\nu,
\end{eqnarray}
and it implies $\tilde s^2_T=s^2_T$.

The open spinor indices $(ij)$ of the TMD operator in eq.~(\ref{def:TMDPDF_Qop}) are to be contracted with different gamma-matrices, which we denote generically as $\Gamma$. The gamma-matrices that appear at the leading order of TMD factorization are
\begin{eqnarray}\label{def:Gamma}
\Gamma=\{\gamma^+,\gamma^+\gamma_5,i\sigma^{\alpha+}_T\gamma_5\},
\end{eqnarray}
where $\sigma^{\alpha +}_T=g^{\alpha\beta}_T \sigma_{\beta\gamma}n^\gamma$, and
\begin{eqnarray}
\sigma^{\mu\nu}=\frac{i}{2}(\gamma^\mu\gamma^\nu-\gamma^\nu\gamma^\mu),\qquad \gamma_5=i\gamma^0\gamma^1\gamma^2\gamma^3=\frac{i}{4!}\epsilon_{\mu\nu\alpha\beta}\gamma^\mu\gamma^\nu\gamma^\alpha\gamma^\beta.
\end{eqnarray}
In the naive parton model interpretation, these gamma-structures are related to the observation of unpolarized ($\gamma^+$), longitudinally polarized ($\gamma^+\gamma^5$) and transversely polarized ($i\sigma^{\alpha+}_T\gamma^5$) quarks inside the hadron. Beyond the leading order factorization one expects that the power suppressed terms of TMD show also different gamma structures. However, currently, the TMD factorization theorem is not established beyond the leading order. Moreover, it is known that TMD distributions with a gamma-structure different from (\ref{def:Gamma}) contain rapidity divergences  that are not renormalized by the standard TMD soft factor~\cite{Gutierrez-Reyes:2017glx}.

Historically, the TMD distributions have been introduced and parameterized in  momentum space~\cite{Bacchetta:2008xw}. Denoting
\begin{eqnarray}
\Phi^{[\Gamma]}_{q\ot h}&=& \frac{1}{2}\Tr\(\Phi_{q\ot h}\Gamma\),
\end{eqnarray}
we have~\cite{Goeke:2005hb,Bacchetta:2006tn} 
\begin{eqnarray}\label{def:gammaP_momentum}
\Phi_{q\ot h}^{[\gamma^+]}(x,p_T)&=& f_1(x,p_T)-\frac{\epsilon_T^{\mu\nu}p_{T\mu}s_{T\nu}}{M}f_{1T}^\perp(x,p_T),
\\\label{def:gamma5_momentum}
\Phi_{q\ot h}^{[\gamma^+\gamma_5]}(x,p_T)&=&
\lambda \,g_{1L}(x,p_T)-\frac{p_{T\mu}s_T^\mu}{M}g_{1T}(x,p_T),
\\\label{def:sigma_momentum}\nn
\Phi^{[i\sigma^{\alpha+}\gamma_5]}_{q\ot h}(x,p_T)&=&s_T^\alpha h_1(x,p_T)+\lambda\frac{p_T^\alpha}{M}h_{1L}^\perp(x,p_T)
\\ &&-\frac{\epsilon_T^{\alpha\mu}p_{T\mu}}{M}h_1^\perp(x,p_T)+\frac{p_T^2}{M^2}\(\frac{g_T^{\alpha\mu}}{2}-\frac{p_T^{\alpha}p_T^{\mu}}{p_T^2}\)s_{T\mu}h_{1T}^\perp (x,p_T),
\end{eqnarray}
where $p_T^2=-\vec p_T^2<0$. Note, that the functions $f(x,p_T)$ depend only on the modulus of $p_T$, but not on the direction. The functions presented here are traditionally called  unpolarized ($f_1$), Sivers ($f_{1T}^\perp$), helicity ($g_{1L}$), worm-gear T ($g_{1T}$), transversity $(h_1)$,
worm-gear L ($h_{1L}^\perp$), Boer-Mulders ($h_1^\perp$) and pretzelosity ($h_{1T}^\perp$) distributions. 

For practical calculations  it is convenient to write TMD distributions in the momentum space as Fourier transform of distributions in the position space
in the  usual manner
\begin{eqnarray}\label{def:p<->b}
\Phi_{q\ot h,ij}(x,p_T)=\int \frac{d^2 \vec b}{(2\pi)^2}e^{+i(\vec b\vec p_T)}\Phi_{q\ot h,ij}(x,\vec b),
\end{eqnarray}
where the scalar product $(\vec b\vec p_T)$ is Euclidian. The decomposition in eq.~(\ref{def:gammaP_momentum}-\ref{def:sigma_momentum}) is then replaced by its analog it position space,
\begin{eqnarray}\label{param:TMDv}
\Phi_{q\ot h}^{[\gamma^+]}(x,\vec b)&=&f_1(x,\vec b)+i\epsilon_T^{\mu\nu} b_\mu s_{T\nu} M f_{1T}^\perp(x,\vec b),
\\\label{param:TMDa}
\Phi^{[\gamma^+\gamma_5]}_{q\ot h}(x,\vec b)&=&\lambda g_{1L}(x,\vec b)+i b_{\mu}s^{\mu}_T M g_{1T}(x,\vec b),
\\\label{param:TMDt}\nn
\Phi^{[i\sigma^{\alpha+}\gamma_5]}_{q\ot h}(x,\vec b)&=&s_{T}^\alpha h_1(x,\vec b)
 -i\lambda b^{\alpha}M h_{1L}^\perp(x,\vec b)
 \\ &&+i\epsilon_{T}^{\alpha\mu}b_\mu M h_1^\perp(x,\vec b)+\frac{M^2\vec b^2}{2}\(\frac{ g_{T}^{\alpha\mu}}{2}+\frac{b^\alpha b^\mu}{\vec b^2}\)s_{T\mu} h_{1T}^\perp(x,\vec b).
\end{eqnarray}
This parameterization coincides\footnote{Comparing parameterization one should take into account that the TMD operator in ref.~\cite{Boer:2011xd} is taken with the vector $b$ oriented in the opposite direction.} with the parameterization given in~\cite{Boer:2011xd}. The explicit transformation rules for all these functions can be found in  appendix \ref{app:p<->b}.

\section{Light-cone expansion for TMD operator}
\label{sec:TMDexpansion}

The small-$b$ matching of TMD distribution to the integrated distributions is obtained by the operator product expansion (OPE) at small-$b$. The OPE is independent from the hadronic states and for this reason it is universal. Let us introduce a separate notation for the TMD operators. The operator that produces TMD distributions in the Drell-Yan case is
\begin{eqnarray}\label{def:TMDop_DY}
\mathcal{U}_{\text{DY}}^{\Gamma}(z,\vec b)&=&\bar q(z n+\vec b)[z n+\vec b,-\infty n+\vec b]\Gamma
[-\infty n-\vec b,-z n-\vec b] q(-zn -\vec b),
\end{eqnarray}
where $\Gamma$ represents the gamma-matrices of the leading set (\ref{def:Gamma}), and the half-infinite Wilson lines are defined as
\begin{eqnarray}
[a_1 n+\vec b,a_2n+\vec b]&=& P\exp\(ig\int_{a_2}^{a_1} d\sigma n^\mu A_\mu(\sigma n+\vec b)\).
\end{eqnarray}
Here and in the following we also omit the T-ordering of the fields, since it is irrelevant in the tree order approximation. The operator that produces the TMD distributions for  SIDIS reads
\begin{eqnarray}\label{def:TMDop_DIS}
\mathcal{U}_{\text{DIS}}^{\Gamma}(x,\vec b)&=&\bar q(z n+\vec b)[z n+\vec b,+\infty n+\vec b]\Gamma
[+\infty n-\vec b,-z n-\vec b] q(-zn -\vec b).
\end{eqnarray}
Generally, the links which connect the end points of Wilson lines at  a distant transverse plane must be added in both operators (for DY and for SIDIS). Here, we omit them for simplicity, assuming that some non-singular gauge (e.g. covariant gauge) is in use. In non-singular gauges the field nullifies at infinities, $A_\mu(\pm \infty n)=0$, and the contribution of distant gauge links vanish. 

The relation between the TMD distribution (\ref{def:TMDPDF_Qop}) and the TMD operator (\ref{def:TMDop_DY}) is
\begin{eqnarray}\label{def:Fourier_to_TMD}
\Phi^{[\Gamma]}_{q\ot h}(x,\vec b)=\int\frac{dz}{2\pi}e^{-2ixzp^+}\langle P,S|\mathcal{U}^\Gamma\(z,\frac{\vec b}{2}\)|P,S\rangle.
\end{eqnarray}

The light-cone expansion of the TMD operators corresponds to the expansion in the variable $\vec b$. The OPE has a generic schematic form
\begin{eqnarray}
\mathcal{U}(z,\vec b)=\sum_i \[C_i\ast \mathcal{O}^{\text{tw2}}_i\](z)+\vec b^\mu \sum_i \[\tilde C_i\ast \mathcal{O}^{\text{tw3}}_{\mu,i}\](z)+O(\vec b^2),
\end{eqnarray}
where $C$'s are Wilson coefficient functions which depend on $\ln \vec b^2$, $\mathcal{O}$ 's are light-cone operators, and the symbol $\ast$ denotes some  integral convolution between coefficient function and operators. Here, the superscripts tw2 and tw3 indicate the \textit{collinear twist}, which in principle differs from the \textit{geometrical twist}. We remind that the term \textit{collinear twist} indicates the distributions which enter the same order of momentum expansion. It is not a well-defined quantum number, in contrast to the \textit{geometrical twist}. The later is defined by "dimension-spin" value, and is a well-defined quantum number, in the sense that e.g. it conserves and does not mix under the scaling transformations. As we will see the operators $\mathcal{O}^{\text{tw3}}$ are compositions of geometrical twist-2 and twist-3 operators. The coefficient functions are perturbatively calculable. In this work, we study the matching only at order  $\alpha_s^0$. 

At leading order in $\alpha_s$ the quantum fields can be considered as classical fields, that satisfy QCD equations of motion. In this approximation, the small-$b$ OPE is just the Taylor expansion at $b=0$.  Expanding  $\mathcal{U}$   in powers of $\vec b$ up to the linear order  we obtain
\begin{eqnarray}\label{U_Taylor}
\mathcal{U}^\Gamma(z,\vec b)=\mathcal{U}^\Gamma(z,\vec 0)+b^\mu \frac{\partial}{\partial  b^\mu} \mathcal{U}^\Gamma(z,\vec b)\Big|_{\vec b=0}+O(\vec b^2).
\end{eqnarray}
The leading term reads
\begin{eqnarray}
\mathcal{U}_{\text{DY}}^{\Gamma}(z,\vec 0)&=&\bar q(z n)[z n,-\infty n]\Gamma [-\infty n,-z n] q(-zn)
=\bar q(z n)[z n,-z n]\Gamma q(-z n),
\end{eqnarray}
where the half-infinite segments of Wilson line compensate each other due to the unitarity of a Wilson line. The same holds for the SIDIS operator
\begin{eqnarray}
\mathcal{U}_{\text{DIS}}^{\Gamma}(z,\vec 0)&=&\bar q(z n)[z n,+\infty n]\Gamma [+\infty n,-z n] q(-zn)=\bar q(z n)[z n,-z n]\Gamma q(-zn).
\end{eqnarray}
Therefore, we obtain that $\mathcal{U}_{\text{DY}}^{\Gamma}(z,\vec 0)=\mathcal{U}_{\text{DIS}}^{\Gamma}(z,\vec 0)$, which is well known.

The term linear in $\vec b^\mu$  is given by the derivative of the operator. Explicitly, it reads
\begin{eqnarray}\label{U_DY_der_0}
\frac{\partial}{\partial b^\mu} \mathcal{U}^{\text{DY}}_\Gamma(z,\vec b)\Big|_{\vec b=0}&=&
\bar q(z n)[z n,-\infty n](\overleftarrow{\partial_{T\mu}}-\overrightarrow{\partial_{T\mu}})\Gamma [-\infty n,-z \lambda] q(-z),
\end{eqnarray}
where $\partial_{T\mu}$ is the derivative with respect to the transverse components only. This expression can be written as
\begin{eqnarray}\label{U_DY_der_1}
\frac{\partial}{\partial b^\mu} \mathcal{U}_{\text{DY}}^\Gamma(z,\vec b)\Big|_{\vec b=0}&=&
\bar q(zn)\overleftarrow{D_\mu}[zn,-\infty n]\Gamma[-\infty n,-zn]q(-zn)
\\\nn &&+ig\int_{-\infty}^z 
\bar q(zn)[zn,\tau n]F_{\mu+}(\tau n)[\tau n,-\infty n]\Gamma[-\infty n,-zn]q(-zn)
\\\nn &&-ig\int^{-\infty}_{-z}
\bar q(zn)[zn,-\infty n]\Gamma[-\infty n,\tau n]F_{\mu+}(\tau n)[\tau n,-zn]q(-zn)
\\\nn &&
-\bar q(zn)[zn,-\infty n]\Gamma[-\infty n,-zn]\overrightarrow{D_\mu}q(-zn),
\end{eqnarray}
where the covariant derivative and the field-strength tensor are defined as usual
\begin{eqnarray}
\overrightarrow{D}_\mu=\overrightarrow{\partial}_\mu-igA_\mu,\qquad
\overleftarrow{D}_\mu=\overleftarrow{\partial}_\mu+igA_\mu,\qquad F_{\mu\nu}=\partial_\mu A_\nu-\partial_\nu A_\mu-ig[A_\mu,A_\nu].
\end{eqnarray}
To obtain the expression (\ref{U_DY_der_1}) we have used  the assumption that\footnote{In singular gauges, one generally cannot expect the boundary condition $A(\pm\infty n)=0$, but $A(\pm \infty n,\vec \infty)=0$. In this case the TMD operator receives the transverse link to corresponding infinity, which preserves the gauge invariance  (for a discussion on the role of singular gauge see
 f.i.~\cite{Belitsky:2002sm,Idilbi:2010im,Idilbi:2010tc,GarciaEchevarria:2011md}). 
Therefore, transverse derivative operator $(\overleftarrow{\partial_T}-\overrightarrow{\partial_T})$ is inserted at the far-end of Wilson lines at $\pm n\infty+\vec \infty$ (compare to (\ref{U_DY_der_1})), and as a result it also differentiates transverse links. Then the expansion formula (\ref{U_DY_der}) obtains an extra term
\begin{eqnarray}\nn
\frac{\partial}{\partial b^\mu} \mathcal{U}_{\text{DY}}^\Gamma(z,\vec b)\Big|_{\vec b=0}&=&\text{(\ref{U_DY_der})}-
2ig
\lim_{b\to 0}\Big\{
\int_{0}^b d\vec \tau^\nu~
\bar q(zn)[zn,-\infty n][-\infty n,-\infty n +\vec \tau]\\\nn &&
\qquad\qquad \Gamma F_{\nu\mu}(-\infty n +\vec \tau)[\vec \tau-\infty n,-\infty n][-\infty n,-zn]q(-zn)\Big\},
\end{eqnarray}
where $\vec \tau=\tau\vec b/|b| $. The limit $b\to 0$ is smooth and thus produces zero. In this way, the result in a singular gauge coincides with the result in a regular gauge. The similar consideration holds for SIDIS operators with replacement $-\infty n\to +\infty n$.} $A(-\infty n)=0$, and the explicit expression for the total derivative of a Wilson line,
\begin{eqnarray}\label{WL_totalDer}
\partial_\mu \{[z_1 n,z_2 n]\}&=&\frac{d}{dy^\mu}[z_1n+y,z_2n+y]\Big|_{y=0}
\\\nn &=&ig\(A_\mu(z_1n)[z_1n,z_2n]-[z_1n,z_2n]A_\mu(z_2n)+\int_{z_2}^{z_1}d\tau [z_1n,\tau n]F_{\mu+}(\tau n)[\tau n,z_2 n]\),
\end{eqnarray}
where the vector $n$ can be arbitrary.

The segments of Wilson line between $-\infty$ and $\tau$ cancel and we obtain
\begin{eqnarray}\label{U_DY_der}
\frac{\partial}{\partial b^\mu} \mathcal{U}_{\text{DY}}^\Gamma(z,\vec b)\Big|_{\vec b=0}&=&
\bar q(zn)\(\overleftarrow{D_\mu}[zn,-zn]-[zn,-zn]\overrightarrow{D_\mu}\)\Gamma q(-zn)
\\\nn &&+ig\(\int_{-\infty}^z +\int_{-\infty}^{-z}\)d\tau~
\bar q(zn)[zn,\tau n]\Gamma F_{\mu+}(\tau n)[\tau n,-zn]q(-zn).
\end{eqnarray}
In the case of SIDIS kinematics the Wilson lines point the future light-like infinity, and therefore, the same derivation gives
\begin{eqnarray}\label{U_DIS_der}
\frac{\partial}{\partial b^\mu} \mathcal{U}_{\text{DIS}}^\Gamma(z,\vec b)\Big|_{\vec b=0}&=&
\bar q(zn)\(\overleftarrow{D_\mu}[zn,-zn]-[zn,-zn]\overrightarrow{D_\mu}\)\Gamma q(-zn)
\\\nn &&-ig\(\int^{\infty}_z +\int^{\infty}_{-z}\)d\tau~
\bar q(zn)[zn,\tau n]\Gamma F_{\mu+}(\tau n)[\tau n,-zn]q(-zn).
\end{eqnarray}
Comparing the results for DY in eq.~(\ref{U_DY_der}) and SIDIS in eq.~(\ref{U_DIS_der}) kinematics we observe that the first term is the same, while the last terms  differ because of the limits of integration and a common sign. Therefore, already at this stage it is clear that the operator in the first term does not contribute to P-odd distributions (i.e. Sivers and Boer-Mulders functions) which is known to change sign in different kinematics.

As expected, the non-compact (in the sense that it spans an infinite range in the position space) TMD operator is expressed via a set of compact light-cone operators. The light-cone operators  in eq.~(\ref{U_DY_der},~\ref{U_DIS_der}) are not very well defined, in the sense, that they are of indefinite \textit{geometrical twist} (more specifically, this issue concerns the first terms of eq.~(\ref{U_DY_der},~\ref{U_DIS_der})). 
At the next stage of the OPE we need to classify the contributions with respect to twist and decompose over independent components. These components are parameterized via parton distributions functions, which are universal and can be measured in different experiment.

 As the key point here is the twist-expansion we provide  some additional discussion. The standard approach to twist-decomposition of operators is to consider their local expansion. In the local expansion the contributions of different twists can be separated by the permutation algebra, and summed back to a non-local representation, see e.g. the detailed decomposition of similar operators in \cite{Belitsky:2000vx}. However, a much simpler approach consists in taking the operator directly in a non-local form \cite{Balitsky:1987bk,Ball:1998sk}. In this approach, one starts with  operators off the light-cone, and makes the twist-decomposition, and then perform the limit to the light-cone. 

In principle, the procedure of  twist-decomposition can be made at the level of operators, see e.g.~\cite{Balitsky:1987bk}. However, practically it is involved, especially for tensor gamma-structure. The evaluation is significantly simpler in the terms of distributions,  e.g. as it is done in ref.~\cite{Ball:1998sk}. Here we are going to follow this second approach. In fact, the derivation presented in the next sections closely follows the procedure described in details in~\cite{Ball:1998sk} for the case of meson distribution amplitudes. The difference in kinematics does not allow us to use the powerful method of conformal basis, but there is no principle difference in other aspects.

Prior to the parameterization and twist-decomposition let us prepare the operator for this procedure, and make its off-light-cone generalization. 
At our order of accuracy (twist-3) we do not need the most general form of  the three-point operators, since they are already of \textit{geometrical twist}-3 and do not contain admixture with twist-2 operators. Therefore, the generalization should be done only for the two-point operators, and it can be simply achieved by the replacement $z n^\mu \to y^\mu$ with $y^2\neq 0$. The result is conveniently re-written in the following form
\begin{eqnarray}\label{der:f1}
\bar q(y)\(\overleftarrow{D_\mu}[y,-y]-[y,-y]\overrightarrow{D_\mu}\)\Gamma q(-y)&=&
\frac{\partial}{\partial y^\mu} \bar q(y)[y,-y]\Gamma q(-y)
\\\nn && \qquad\qquad-ig \int_{-1}^1 dv \, v y^\nu \bar q(y)[y,vy]\Gamma F_{\mu\nu}(v y)[vy,-y]q(-y),
\end{eqnarray}
where we have used the formula for the stretch derivative of the Wilson line
\begin{eqnarray}\label{WL_xDer}
\frac{\partial}{\partial y^\mu} [y,-y]&=&ig\(A_\mu(y)[y,-y]+[y,-y]A_\mu(-y)+\int_{-1}^1dv v y^\nu [y,v y]F_{\mu\nu}(v y)[v y,-y]\).
\end{eqnarray}
Note, that this expression is the same for DY and SIDIS operators. The last term of (\ref{der:f1}) is again pure twist-3 operator, and thus one can set it directly on the light-cone.

Let us conclude this section with an intermediate summary of our main results. For convenience we introduce the generic notation for two- and three-point operators
\begin{eqnarray}
\label{def:O}
\mathcal{O}_{\Gamma}(z)&=&\bar q(z n)[z n,-z n]\Gamma q(-zn),
\\
\label{def:T}
\mathcal{T}_\Gamma^\mu(z_1,z_2,z_3)&=&g\bar q(z_1 n)[z_1 n, z_2 n]\Gamma F^{\mu+}(z_2 n)[z_2 n,z_3 n]q(z_3 n).
\end{eqnarray}
The expression for the first terms of small-$b$ expansion for TMD operator reads (at leading order in $\alpha_s$ )
\begin{eqnarray}\label{U_DY_GAMMA}
\mathcal{U}_{\text{DY}}^\Gamma(z,\vec b)&=&\mathcal{O}_\Gamma(z)+b_\mu\Big\{\lim_{y\to zn}\frac{\partial}{\partial y_\mu}\mathcal{O}_\Gamma(y)-i\int_{-1}^1 dv \,v z \,\mathcal{T}_\Gamma^\mu(z,vz,-z)
\\\nn &&\qquad\qquad\qquad\qquad +i\(\int_{-\infty}^z+\int_{-\infty}^{-z}\)d\tau \mathcal{T}_\Gamma^\mu(z,\tau,-z)\Big\}
 +O(\vec b^2),
\end{eqnarray}
\begin{eqnarray}\label{U_DIS_GAMMA}
\mathcal{U}_{\text{DIS}}^\Gamma(z,\vec b)&=&\mathcal{O}_\Gamma(z)+ b_\mu\Big\{\lim_{y\to zn}\frac{\partial}{\partial y_\mu}\mathcal{O}_\Gamma(y)-i\int_{-1}^1 dv \,vz\,\mathcal{T}_\Gamma^\mu(z,vz,-z)
\\\nn &&\qquad\qquad\qquad\qquad -i\(\int^{\infty}_z+\int^{\infty}_{-z}\)d\tau \mathcal{T}_\Gamma^\mu(z,\tau,-z)\Big\}
 +O(\vec b^2).
\end{eqnarray}
The limit $y\to z n$ implies $y^2\to 0$ such that the light-like separation between fields is $z$.

\section{Collinear distributions}
\label{sec:collinear}

Evaluating the matrix elements of eq.~(\ref{U_DY_GAMMA},~\ref{U_DIS_GAMMA}), and hence the matching of TMD distributions,
 we meet with a number of collinear parton distributions. In this section we present the parameterization of two and three point parton distributions that appear in the final result. In fact, the  functions that we find represent a complete set of \textit{geometrical twist} 2 and 3 quark distributions. For the two-point functions we use the standard parameterization by \cite{Jaffe:1991ra}. For the three point functions there is not a commonly accepted parameterization, therefore, we introduce a parameterization inspiring  in \cite{Braun:2009mi}.

\subsection{Parameterization of quark-quark correlators}

The standard parameterization of light-cone quark-quark correlators is given \cite{Jaffe:1991ra} and reads
\begin{eqnarray}\label{param:vPDF}
\langle P,S|O^{\gamma^\mu}(z)|P,S\rangle&=&2\int dx e^{2ix z p^+ }\Big\{p^\mu f_1(x)+\frac{n^\mu}{(np)}M^2 f_4(x)\Big\},
\\\label{param:aPDF}
\langle P,S|O^{\gamma^\mu\gamma^5}(z)|P,S\rangle&=&2\int dx e^{2ixzp^+}\Big\{\lambda p^\mu g_1(x)+s_T^\mu M g_T(x)+\lambda  M^2 \frac{n^\mu}{(np)}g_3(x)\Big\},
\\\label{param:tPDF}
\langle P,S|O^{i\sigma^{\mu\nu}\gamma^5}(z)|P,S\rangle&=& 2\int dx e^{2ixzp^+}\Big\{(s_T^\mu p^\nu-p^\mu s_T^\nu) h_1(x)+\lambda \frac{M}{(np)}(p^\mu n^\nu-n^\mu p^\nu)h_L(x)
\\\nn &&\qquad\qquad\qquad\qquad\qquad\qquad\qquad+(s_T^\mu n^\nu-n^\mu s_T^\nu)\frac{M^2}{(np)} h_3(x)\Big\},
\end{eqnarray}
where the operators $\mathcal{O}^\Gamma$ are defined in eq.~(\ref{def:O}). The twist-2 PDFs $f_1$, $g_1$ and $h_1$ are known as unpolarized, helicity and transversity PDFs. The PDFs $g_T$ and $h_L$ are of \textit{collinear} twist-3. The PDFs $f_4$, $g_3$ and $h_3$ are of \textit{collinear} twist-4, and do not appear in the current  final results.  The \textit{collinear} twist-3 PDFs are not independent as they are combinations of PDFs of twist-2 and three-point PDFs. The derivation of this relation can be done with the help of QCD equations of motion and is presented in the appendix \ref{app:EOM}.

The PDF defined by eq.~(\ref{param:vPDF},~\ref{param:aPDF},~\ref{param:tPDF}) are non-zero for $-1<x<1$  and zero for $|x|>1$ \cite{Jaffe:1983hp}. They can be represented by
\begin{eqnarray}
f_1(x)=\theta(x)q(x)-\theta(-x)\bar q(x),
\end{eqnarray}
where $q(x)$ and $\bar q(x)$ are the usual quark and anti-quark parton densities in the infinite momentum frame. A similar interpretation holds for $g_1$ and $h_1$.

At $z\to 0$ the operators turn to local operators. The matrix elements of local operator can be parameterized in terms of the corresponding charges. This implies the existence of exact relations relations among the  first moments of PDFs. 
In the present case the important relations are
\begin{eqnarray}\label{firstmoments}
\int_{-1}^1 dx g_1(x)=\int_{-1}^1 dx g_T(x),\qquad \int_{-1}^1 dx h_1(x)=\int_{-1}^1 dx h_L(x),
\end{eqnarray}
and they  are another form of the Burkhard-Cottingham sum rule \cite{Burkhardt:1970ti}.

In order to proceed with the matching, we need also a parameterization  of off light-cone collinear functions.
In general, the parameterization of matrix elements off light-cone does not coincide with the parameterization of light-cone matrix elements, which is given in eq.~(\ref{param:vPDF},~\ref{param:aPDF},~\ref{param:tPDF}). 
However, on and off light-cone parameterizations
can be related to each other order by order in the expansion over $y^2$ (where $y$ is the distance between quark fields), see e.g. discussion in~\cite{Ball:1998ff}. 
Such relations up to linear terms in $y$ are presented in appendix \ref{app:offLight}.
 Using the off-light-cone parameterization of eq.~(\ref{app:vPDF},~\ref{app:aPDF},~\ref{app:tPDF}) we derive the matrix elements of the first terms in the small-$b$ OPE in eq.~(\ref{U_DY_GAMMA},~\ref{U_DIS_GAMMA}). We find
\begin{eqnarray}\label{derO:v}
n_\alpha g_T^{\mu\nu}\lim_{y\to zn}\frac{\partial}{\partial y^\nu}\langle P,S|O^{\gamma^\alpha}(y)|P,S\rangle&=&0,
\\\label{derO:a}
n_\alpha g_T^{\mu\nu}\lim_{y\to zn}\frac{\partial}{\partial x^\nu}\langle P,S|O^{\gamma^\alpha\gamma^5}(y)|P,S\rangle&=&
2s_T^\mu M\int du e^{2ix zp^+} \frac{g_1(x)-g_T(x)}{z},
\\\label{derO:t}
n_\gamma g_T^{\alpha \beta} g_T^{\mu\nu}\lim_{y\to zn}\frac{\partial}{\partial y^\nu}\langle P,S|O^{i\sigma^{\beta\gamma}\gamma^5}(y)|P,S\rangle&=&
2\lambda M g_T^{\mu\alpha}\int dx e^{2ix zp^+} \frac{h_1(x)-h_L(x)}{z}.
\end{eqnarray}
Moreover these expressions depend on the particular off-light-cone parameterization that is used. 
In any case, the functions $g_T$ and $h_L$ are not independent, and must be expressed in  terms of distributions with definite \textit{geometrical twist}. Such a re-expression is also dependent on the parameterization. In the final result all (intermediate and off-light-cone) parameterization dependence cancels, and the result is uniquely defined using definite twist distributions.

\subsection{Parameterization of quark-gluon-quark correlators}

The parameterization of matrix elements of a three-point operator has the following general structure
\begin{eqnarray}\label{general_parameterization}
\langle P,S|\mathcal{T}^\mu_\Gamma (z_1,z_2,z_3)|P,S\rangle&=&\sum_{i}t_\Gamma^{i;\mu...}(P,S,n,g_T,\epsilon_T)\int [dx]e^{-ip^+(x_1z_1+x_2z_2+x_3z_3)}F(x_1,x_2,x_3),
\end{eqnarray}
where the integration measure is \cite{Jaffe:1983hp}
\begin{eqnarray}\label{def:[dx]}
[dx]=dx_1dx_2dx_3\delta(x_1+x_2+x_3),\qquad -1<x_{1},x_2,x_3<1.
\end{eqnarray}
In the rest of the paper we use the tilde notation for Fourier images of the functions
\begin{eqnarray}\label{def:Fourier}
\tilde F(z_1,z_2,z_3)=\int [dx]e^{-ip^+(x_1z_1+x_2z_2+x_3z_3)}F(x_1,x_2,x_3)\ .
\end{eqnarray}
In
eq.~(\ref{general_parameterization})
we  have introduced a tensor $t$  built of $P^\mu$, (single entry of)$S^\mu$, $n^\mu$, $g_T^{\mu\nu}$ and $\epsilon_T^{\mu\nu}$, and their scalar products, such that it preserves the permutation symmetry of indices on left-hand side, and it is invariant under rescaling $z\to \alpha z$. 
Such a tensor contains significant number of terms, which can be restricted by discrete symmetries, such as parity, time-reversal and charge-conjugation (which can be replaced by hermiticity due to CPT theorem). 
The parity invariance results into a relation among the terms of eq.~(\ref{general_parameterization})
\begin{eqnarray}\label{P-tranform}
t_\Gamma^{i;\mu...}(P,S,n,g_T,\epsilon_T)F(x_1,x_2,x_3)=
\eta_{P}^{\Gamma;\mu...} t_\Gamma^{i;\mu...}(\bar P,s_T,\bar  n,g_T,-\epsilon_T)F(x_1,x_2,x_3),
\end{eqnarray}
where the bar denotes the parity transformation of a vector $\bar v^\mu=v_\mu$, and $\eta_P^{\Gamma,\mu...}$ is the sign factor that appears in the parity transformation of the operator $\mathcal{P}\mathcal{T}_\Gamma \mathcal{P}^\dagger =\eta_P^\Gamma \mathcal{T}_\Gamma$. The time reversal transformation results into
\begin{eqnarray}\label{T-tranform}
t_\Gamma^{i;\mu...}(P,S,n,g_T,\epsilon_T)F(x_1,x_2,x_3)=
\eta_{T}^{\Gamma;\mu...} t_\Gamma^{i;\mu...}(\bar P,-s_T,-\bar n,-g_T,-\epsilon_T)F(-x_3,-x_2,-x_1),
\end{eqnarray}
where $\eta_T^{\Gamma,\mu...}$ is the sign factor that appears in the time-reversal transformation of the operator $T\mathcal{T}_\Gamma T^\dagger =\eta_T^\Gamma \mathcal{T}_\Gamma$. In contrast to the two-point functions the time-reversal symmetry does not restrict the number of tensor structures $t_i$, because the functions on left- and right-hand sides of eq.~(\ref{T-tranform}) are of different arguments. Additionally one has the hermiticity relation which gives
\begin{eqnarray}\label{H-tranform}
\eta^\Gamma_H F^*(-x_1,-x_2,-x_3)=F(x_3,x_2,x_1),
\end{eqnarray}
where $\eta_H$ is sign of hermitian conjugation of the operator $(\mathcal{T}_\Gamma)^\dagger=\eta_H^\Gamma \mathcal{T}_\Gamma$ (here we expect that the tensors $t$ are real). Together the time-reversal (\ref{T-tranform}) and hermiticity (\ref{H-tranform}) relations dictates the complex and symmetry properties of the functions $F$.

In general the number of tensors $t$ is very large. However, for the current work we need only the tensors which are non-zero if open indices are transverse, and the rest of indices are contracted with $n^\mu$. In other words, we require the tensor structure of collinear twist-3. We  find four such functions 
\begin{eqnarray}\label{def:PDF_T}
\langle P,S|\mathcal{T}^\mu_{\gamma^+}|P,S\rangle&=&2(p^+)^2 \tilde{s}^\mu_T M \int [dx]e^{-ip^+(x_1z_1+x_2z_2+x_3z_3)}T(x_1,x_2,x_3),
\\\label{def:PDF_DeltaT}
\langle P,S|\mathcal{T}^\mu_{\gamma^+\gamma^5}|P,S\rangle&=&2i(p^+)^2 s^\mu_T M \int [dx]e^{-ip^+(x_1z_1+x_2z_2+x_3z_3)}\Delta T(x_1,x_2,x_3),
\\\label{def:PDF_deltaT}
\langle P,S|\mathcal{T}^\mu_{i\sigma^{\alpha+}\gamma^5}|P,S\rangle&=&2(p^+)^2 \epsilon_T^{\mu\alpha} M \int [dx]e^{-ip^+(x_1z_1+x_2z_2+x_3z_3)}\delta T_\epsilon(x_1,x_2,x_3)
\\\nn &&  +2i(p^+)^2 \lambda g_T^{\mu\alpha} M \int [dx]e^{-ip^+(x_1z_1+x_2z_2+x_3z_3)}\delta T_g(x_1,x_2,x_3).
\end{eqnarray}
Here, the factors $M$ are set to have dimensionless three-point PDFs $T$. The definition of distributions $T$ and $\Delta T$ coincides\footnote{To compare with ref.\cite{Braun:2009mi}, we note that their definition of $\tilde s$ has opposite to us sign. Also during comparison we facilitate $s^2=-1$.} with the definition used in \cite{Braun:2009mi}, up to a factor $M$. The comparison to 
ETQS\footnote{ETQS is acronym for Efremov-Teryaev-Qiu-Sterman \cite{Efremov:1983eb,Qiu:1991pp}.}
 functions (here we compare to definitions in eq.~(12) and eq.~(21) of \cite{Kang:2008ey}) gives
\begin{eqnarray}\label{T<->QS}
\widetilde{\mathcal{T}}_{q,F}(x,x+x_2)= M T(-x-x_2,x_2,x),\qquad \widetilde{\mathcal{T}}_{\Delta q,F}(x,x+x_2)= M \Delta T(-x-x_2,x_2,x).
\end{eqnarray}

The distribution $T$ are real dimensionless functions. According to eq.~(\ref{T-tranform}) they obey the following symmetry properties
\begin{eqnarray}\label{sym_prop1}
T(x_1,x_2,x_3)&=&T(-x_3,-x_2,-x_1),
\\\label{sym_prop2}
\Delta T(x_1,x_2,x_3)&=&-\Delta T(-x_3,-x_2,-x_1),
\\\label{sym_prop3}
\delta T_\epsilon(x_1,x_2,x_3)&=&\delta T_\epsilon(-x_3,-x_2,-x_1),
\\\label{sym_prop4}
\delta T_g(x_1,x_2,x_3)&=&-\delta T_g(-x_3,-x_2,-x_1).
\end{eqnarray}
The Fourier transform of these distributions obey the same symmetry properties. These four functions are the only genuine twist-3 distributions in the quark sector.

It appears very convenient to introduce the following integral combinations,
\begin{eqnarray}\label{def:Tn}
T^{(n)}(x)&=&\int \frac{[dx]}{x_2^n}\(\delta(x-x_3)+(-1)^n \delta(x+x_1)\)T(x_1,x_2,x_3),
\\\label{def:DeltaTn}
\Delta T^{(n)}(x)&=&\int \frac{[dx]}{x_2^n}\(\delta(x-x_3)-(-1)^n \delta(x+x_1)\)\Delta T(x_1,x_2,x_3),
\\\label{def:deltaTne}
\delta T_\epsilon^{(n)}(x)&=&\int \frac{[dx]}{x_2^n}\(\delta(x-x_3)+(-1)^n \delta(x+x_1)\)\delta T_\epsilon(x_1,x_2,x_3),
\\\label{def:deltaTng}
\delta T_g^{(n)}(x)&=&\int \frac{[dx]}{x_2^n}\(\delta(x-x_3)-(-1)^n \delta(x+x_1)\)\delta T_g(x_1,x_2,x_3).
\end{eqnarray}
The one-variable functions $T^{(n)}$, $\Delta T^{(n)}$ and $\delta T^{(n)}$ are in some aspects similar to the usual PDFs. For example, they have zero boundary conditions,
\begin{eqnarray}
T^{(n)}(\pm 1)=0, \qquad \Delta T^{(n)}(\pm 1)=0,\qquad \delta T_\epsilon^{(n)}(\pm 1)=0,\qquad \delta T_g^{(n)}(\pm 1)=0.
\end{eqnarray}
In the following, we intensively use the functions in eq.~(\ref{def:Tn}-\ref{def:deltaTng}), since they naturally arise and describe the worm-gear functions and allow a simplification of formulas.

\section{Leading matching of TMD distributions}
\label{sec:matching}

In this section we assemble the result for the leading matching of TMD distributions up to terms linear in $\vec b$. 
For this purpose  we need to evaluate the matrix element of the operators in eq.~(\ref{U_DY_GAMMA},~\ref{U_DIS_GAMMA}) using the parameterizations introduced in the previous section. Here we should take into account the decomposition of \textit{collinear} twist-3 distributions over the distributions with definite \textit{geometrical twist}. In the following subsections we consider each gamma-structure individually, and discuss the features of its evaluation. For convenience we also collect the final results in sec.~\ref{sec:results}.

\subsection{Vector operator}

We start with the study of the vector operator, i.e. with $\Gamma=\gamma^+$, in the DY kinematics. Taking the forward matrix element of the operator relation in eq.~(\ref{U_DY_GAMMA}) we obtain
\begin{eqnarray}\label{th:vector1}
\langle P,S|\mathcal{U}_{\text{DY}}^{\gamma^+}(z,\frac{\vec b}{2})|P,S\rangle&=&2p^+ \int dx e^{2ix zp^+}f_1(x)+2(p^+)^2 M \tilde s^\mu \frac{b_\mu}{2} \bigg[
\\&&\nn-i\int_{-1}^1 dv vz \tilde T(z,vz,-z)+i\(\int_{-\infty}^z+\int_{-\infty}^{-z}\)d\tau \tilde T(z,\tau,-z)\bigg]+O(\vec b^2),
\end{eqnarray}
where the contribution of the two-point correlator vanishes in accordance to eq.~(\ref{derO:v}). 

The function $T(z,vz -z)$ is symmetric in $v$ due to the symmetry relation in eq.~(\ref{sym_prop1}). Therefore, the anti-symmetric integral, which is the first in the square brackets of eq.~(\ref{th:vector1}), vanishes,
\begin{eqnarray}\label{th:intvT=0}
\int_{-1}^{1} dv\,vz\,\tilde T(z,vz,-z)&=&0.
\end{eqnarray}
In this way, the contributions linear in $b$  are represented by a single entry, namely, by the last term of eq.~(\ref{th:vector1}). Using the reflection of coordinates in eq.~(\ref{sym_prop1}) we present it as
\begin{eqnarray}\label{th:T=SQ}
\(\int_{-\infty}^z+\int_{-\infty}^{-z}\) d\tau \tilde T(z,\tau,-z)
&=&
\int^{\infty}_{-\infty} d\tau \tilde T(z,\tau,-z).
\end{eqnarray}
Taking into account these simplifications we find
\begin{eqnarray}\label{th:vector2}
\langle P,S|\mathcal{U}_{\text{DY}}^{\gamma^+}(z,\frac{\vec b}{2})|P,S\rangle&=&2p^+ \int dx e^{2ix zp^+}f_1(x)+i(p^+)^2 M \tilde s^\mu b_\mu \int_{-\infty}^{\infty}d\tau \tilde T(z,\tau,-z)+O(\vec b^2).
\end{eqnarray}

In the case of SIDIS kinematic the operators are given by eq.~(\ref{U_DIS_GAMMA}). Applying the same procedure we find
\begin{eqnarray}\label{th:vector3}
\langle P,S|\mathcal{U}_{\text{DIS}}^{\gamma^+}(z,\frac{\vec b}{2})|P,S\rangle&=&2p^+ \int dx e^{2ix zp^+}f_1(x)-i(p^+)^2 M \tilde s^\mu b_\mu \int_{-\infty}^{\infty}d\tau \tilde T(z,\tau,-z)+O(\vec b^2),
\end{eqnarray}
where we have used
\begin{eqnarray}\label{th:T=SQ2}
\(\int^{\infty}_z+\int^{\infty}_{-z}\) d\tau \tilde T(z,\tau,-z)&=&
\int^{\infty}_{-\infty} d\tau \tilde T(z,\tau,-z).
\end{eqnarray}
The only difference between Drell-Yan,  eq.~(\ref{th:vector2}) and SIDIS,  eq.~(\ref{th:vector3}) cases is the sign of the linear term. It corresponds to the famous process dependence of the Sivers function \cite{Collins:2002kn}.

The TMD distribution is obtained by Fourier transformation  over the light-cone distance, 
eq.~(\ref{def:Fourier_to_TMD}). Performing it we obtain
\begin{eqnarray}\label{th:vectorresult1}
\text{(DY)}\qquad \Phi^{[\gamma^+]}_{q\ot h}(x,\vec b)&=& f_1(x)+i b_\mu \tilde s_T^\mu M\, \pi T(-x,0,x)+O(\vec b^2),
\\\label{th:vectorresult2}
\text{(SIDIS)}\qquad \Phi^{[\gamma^+]}_{q\ot h}(x,\vec b)&=& f_1(x)-i b_\mu \tilde s_T^\mu M\, \pi T(-x,0,x)+O(\vec b^2).
\end{eqnarray}
Here we have used,
\begin{eqnarray}
&&\int_{-\infty}^\infty \frac{dz}{2\pi}\int_{-\infty}^\infty d\tau e^{-2ixzp^+}\tilde T(z,\tau,-z)=\frac{\pi}{(p^+)^2}T(-x,0,x).
\end{eqnarray}
These expressions represent the leading matching of vector TMD distribution. Comparing it to the parameterization in eq.~(\ref{param:TMDv}) we find the matching of individual functions. 
Naturally, the unpolarized TMDPDF matches the unpolarized PDF, $f_1(x,\vec b)=f_1(x)+O(\vec b^2)$. The Sivers function matching is process dependent  and it reads
\begin{eqnarray}\label{th:sivers_DY}
\text{(DY)}\qquad f_{1T}^\perp(x,\vec b)&=&\pi T(-x,0,x)+O(\vec b^2),
\\\label{th:sivers_DIS}
\text{(SIDIS)}\qquad f_{1T}^\perp(x,\vec b)&=&-\pi T(-x,0,x)+O(\vec b^2).
\end{eqnarray}
Note, that the correction term is proportional to $\vec b^2$, and therefore, generically, contains twist-5 functions (and twist-4 functions for unpolarized distribution).

These expression, albeit in the different form, are well-known. In the two-point notation for ETQS function (\ref{T<->QS}), the central value of three-point function $T(-x,0,x)$ corresponds to the diagonal value $\widetilde{\mathcal{T}}_{q,F}(x,x)$. Therefore, we can compare (\ref{th:sivers_DY},~\ref{th:sivers_DIS}) to the expressions given in literature, where certain momentum space moments are calculated. Using the transformation rules presented in appendix \ref{app:p<->b}, one can check that 
\begin{eqnarray}
\int d^2\vec p_T \frac{\vec p_T^2}{M^2}f_{1T}^\perp(x,p_T)=2\pi T(-x,0,x),\qquad \int d^2\vec p_T e^{-i (\vec b \vec p_T)} \frac{p_T^\alpha}{M} f_{1T}^\perp(x,p_T)=i\pi b^\alpha T(-x,0,x).
\end{eqnarray}
Here the sign is given for the DY case, and should be changed for the SIDIS case.
To our best understanding\footnote{The comparison can not be done accurately in all cases, since some articles do not provide the full details on sign conventions and definitions.} these expression coincide with ones presented in \cite{Boer:2003cm,Ji:2006ub,Kang:2011mr,Kang:2011hk}.

\subsection{Axial operator}

Taking the forward matrix element of the operator in eq.~(\ref{U_DY_GAMMA}) with $\Gamma=\gamma^+\gamma^5$, we obtain
\begin{eqnarray}
\langle P,S|\mathcal{U}_{\text{DY}}^{\gamma^+\gamma^5}(z,\frac{\vec b}{2})|P,S\rangle&=&2\lambda p^+ \int dx e^{2ix zp^+}g_1(x)+2M s^\mu_T \frac{b_\mu}{2} \bigg[\int du e^{2iuzp^+}\frac{g_1(u)-g_T(u)}{z}\label{eq:ax}
\\&&+(p^+)^2\int_{-1}^1 dv vz \Delta \tilde T(z,vz,-z)-(p^+)^2\(\int_{-\infty}^z+\int_{-\infty}^{-z}\)d\tau \Delta \tilde T(z,\tau,-z)\bigg]+O(\vec b^2),
\nn
\end{eqnarray}
where we have used the parameterizations  in eq.~(\ref{param:aPDF},~\ref{def:PDF_DeltaT}) and the relation of eq.~(\ref{derO:a}). 

To proceed further we take the inverse Fourier transform. We have observed that these integrals naturally enter into the moments of the three-point functions, which are defined in eq.~(\ref{def:Tn}-\ref{def:deltaTng}). Moreover, it is convenient to present  them as a Mellin convolution. Using these tricks we find
\begin{eqnarray}\label{th:axial1}
&&\int \frac{dz}{2\pi}e^{-2ixzp^+} \int_{-1}^1 dv vz \Delta \tilde T(z,vz,-z)=\frac{i}{(p^+)^2}\Big[\frac{\Delta T^{(1)}(x)}{2}+\int_{-1}^1 du \int_0^1 dy \,
u\Delta T^{(2)}(u) \delta(x-yu)\Big],
\\
\label{th:int_g1-gT}
&&\int \frac{dz}{2\pi}e^{-2ix zp^+}\int du e^{2iu zp^+} \frac{g_1(u)-g_T(u)}{z}=i\int_{-1}^{1} du \int_0^1 dy\, u(g_1(u)-g_T(u))\delta(x-uy).
\end{eqnarray}

The last integral in eq.~(\ref{eq:ax}) over the  process-dependent term does not vanish, \begin{eqnarray}
\label{th:axialOP_int1}
&&\int \frac{dz}{2\pi}e^{-2ixzp^+} \(\int_{-\infty}^z+\int_{-\infty}^{-z}\)d\tau \Delta \tilde T(z,\tau,-z)=\frac{i}{(p^+)^2}\frac{\Delta T^{(1)}(x)}{2},
\\\label{th:axialOP_int2}
&&\int \frac{dz}{2\pi}e^{-2ixzp^+} \(\int^{\infty}_z+\int^{\infty}_{-z}\)d\tau \Delta \tilde T(z,\tau,-z)=\frac{-i}{(p^+)^2}\frac{\Delta T^{(1)}(x)}{2},
\end{eqnarray}
where we have used the assumption that the integrand goes to zero at infinity. The sign difference between these integrals, is compensated by the common sign difference in the operators for DY,  eq.~(\ref{U_DY_GAMMA}) and SIDIS, eq.~ (\ref{U_DIS_GAMMA}) kinematics. Therefore, the contribution of seemingly process-dependent terms is the same for both operators. It is exactly compensated by the contribution of eq.~(\ref{th:axial1}), and thus the function $\Delta T^{(1)}$ drops out of calculation.

Combining all together we obtain the same result for DY and SIDIS kinematics, which is
\begin{eqnarray}\label{th:axialresult1}
\Phi^{[\gamma^+\gamma^5]}(x,\vec b) &=&\lambda g_1(x)+i b_\mu s_T^\mu M\int_{-1}^{1} du \int_0^1 dy \delta(x-uy)u\(
g_1(u)-g_T(u)+\Delta T^{(2)}(u)\).
\end{eqnarray}

Comparing to parameterizations in eq.~(\ref{param:TMDa}) we find that the matching for the helicity TMD distribution $g_{1L}(x,\vec b)=g_1(x)+O(\vec b^2)$, and for the worm-gear-T distribution is
\begin{eqnarray}\label{th:g1T-via-gT}
g_{1T}(x,\vec b)=\int_{-1}^{1} du \int_0^1 dy \delta(x-uy)u\(g_1(u)-g_T(u)+\Delta T^{(2)}(u)\)+O(\vec b^2).
\end{eqnarray}
The expression (\ref{th:g1T-via-gT}) is not the final one, because the function $g_T$ can be rewritten via functions of definite twist,
\begin{eqnarray}\label{th:gT_final}
g_T(x)&=&\int_0^1 dy \int_{-1}^1 du \delta(x-yu)\Big[g_1(u) +\frac{T^{(1)}(u)-\Delta T^{(1)}(u)-\varepsilon_+ h_1(u)}{2u}(1-\delta(\bar y))+\Delta T^{(2)}(u)\Big],
\end{eqnarray}
where $\varepsilon_+=2m/M$ with $m$ being the mass of a quark and $\bar y=1-y$. The derivation of this decomposition is given in the appendix \ref{app:gT}. It is straightforward to check that it obeys the Burkhard-Cottingham sum rule eq.~(\ref{firstmoments}). Inserting the function $g_T$ into eq.~(\ref{th:g1T-via-gT}) and using the associativity of Mellin transformation (see also eq.~(\ref{app:gT_convolution})) we obtain 
\begin{eqnarray}\label{th:g1T-via-T}
g_{1T}(x,\vec b)=x\int_{-1}^{1} du \int_0^1 dy \delta(x-uy)\(g_1(u)+\Delta T^{(2)}(u)+\frac{T^{(1)}(u)-\Delta T^{(1)}(u)-\varepsilon_+ h_1(u)}{2u}\)+O(\vec b^2).
\end{eqnarray}
This is the final form of the matching of the worm-gear function to the twist-2 and twist-3 functions. The Mellin convolution, which is presented in eq.~(\ref{th:g1T-via-T}) by $\delta$-function, can be explicitly integrated. It gives the following representation
\begin{eqnarray}\label{th:g1T-via-T(ver2)}
g_{1T}(x,\vec b)=x \int_x^1 \frac{du}{u}\(g_1(u)+\Delta T^{(2)}(u)+\frac{T^{(1)}(u)-\Delta T^{(1)}(u)-\varepsilon_+ h_1(u)}{2u}\)+O(\vec b^2),\qquad \text{for }x>0,
\\
g_{1T}(x,\vec b)=x \int^{x}_{-1} \frac{du}{|u|}\(g_1(u)+\Delta T^{(2)}(u)+\frac{T^{(1)}(u)-\Delta T^{(1)}(u)-\varepsilon_+ h_1(u)}{2u}\)+O(\vec b^2),\qquad \text{for }x<0,
\end{eqnarray}
The obtained result can be compared to the first transverse momentum moments of TMD distribution derived in ref.~\cite{Kanazawa:2015ajw} (see eq.~(47)), and agrees with it..

\subsection{Tensor operator}

The matrix element of the tensor operator in eq.~(\ref{U_DY_GAMMA}) (i.e. with $\Gamma=i\sigma^{\alpha+}_T\gamma^5$) has a more complicated form
\begin{eqnarray}
\langle P,S|\mathcal{U}_{\text{DY}}^{i\sigma^{\alpha+}_T\gamma^5}(z,\frac{\vec b}{2})|P,S\rangle&=&2s_T^\mu p^+ \int dx e^{2ix zp^+}h_1(x)+2M \frac{b_\mu}{2} \bigg[\lambda g_T^{\mu\alpha}\int du e^{2iuzp^+}\frac{h_1(u)-h_L(u)}{z}
\\&&\nn+(p^+)^2\int_{-1}^1 dv vz \(\lambda g_T^{\mu\alpha}\delta \tilde T_g(z,vz,-z)-i\epsilon_T^{\mu\alpha}\delta \tilde T_\epsilon(z,vz,-z)\)
\\\nn && -(p^+)^2\(\int_{-\infty}^z+\int_{-\infty}^{-z}\)d\tau \(\lambda g_T^{\mu\alpha}\delta \tilde T_g(z,\tau,-z)-i\epsilon_T^{\mu\alpha}\delta \tilde T_\epsilon(z,\tau,-z)\)\bigg]+O(\vec b^2),
\end{eqnarray}
where we have used the parameterizations  of eq.~(\ref{param:tPDF},~\ref{def:PDF_deltaT}) and the relation (\ref{derO:t}). Its structure repeats the structure discussed during evaluations of the vector operator (for terms proportional to $\delta T_\epsilon$) and axial operator (for terms proportional to $\delta T_g$). Therefore, we skip the discussion on the Fourier integrals and write down the final expression for the matching of transversally polarized TMD distribution. We obtain (compare to eq.(\ref{th:vectorresult2},~\ref{th:vectorresult1}) and (\ref{th:axialresult1}))
\begin{eqnarray}\label{th:tensorresult1}
\Phi^{[i\sigma^{\alpha+}\gamma^5]}(x,\vec b)&=&s_T^\alpha h_1(x)\pm ib_\mu\epsilon^{\mu\alpha}_T \pi \delta T_\epsilon(-x,0,x)
\\\nn &&+i\lambda b^\alpha M \int_{-1}^1 du \int_0^1 dy \delta (x-uy)u\(h_1(u)-h_L(u)+\delta T_g^{(2)}(u)\)+O(\vec b^2),
\end{eqnarray}
where the upper sign should be taken for the DY kinematics, and lower for the SIDIS kinematics.

Comparing eq.~(\ref{th:tensorresult1}) with the parameterization of eq.~(\ref{param:TMDt}) we obtain the matching of individual TMD distributions. The transversity distribution is $h_1(x,\vec b)=h_1(x)+O(\vec b^2)$. The Boer-Mulders functions depends on the underling process and reads
\begin{eqnarray}
\text{(DY)}\qquad h_1^\perp(x,\vec b)&=&-\pi \delta T_\epsilon(-x,0,x)+O(\vec b^2),
\\
\text{(SIDIS)}\qquad h_1^\perp(x,\vec b)&=&\pi \delta T_\epsilon(-x,0,x)+O(\vec b^2).
\end{eqnarray}
The worm-gear L function is independent on the process and has the expression
\begin{eqnarray}
h_{1L}^\perp(x,\vec b)&=&-\int_{-1}^1 du \int_0^1 dy \delta (x-uy)u\(h_1(u)-h_L(u)+\delta T_g^{(2)}(u)\)+O(\vec b^2).
\end{eqnarray}
The pretzelosity distribution has no matching at this level of accuracy, despite the fact that the matrix element over free quarks is non-zero at $\vec b^2\to 0$ \cite{Gutierrez-Reyes:2018iod}. It is expected that the first non-zero contribution to the pretzelosity is of twist-4.

As in the case of the worm-gear T function, the expression for the worm-gear L function should be rewritten via a definite twist function. The derivation of function $h_L$ is given appendix \ref{app:hL}. It reads
\begin{eqnarray}\label{th:hL_final}
h_L(x)&=&-\int_0^1 dy \int_{-1}^1 du \delta(x-y u)\Big[2y(h_1(u)+\delta T^{(2)}_g(u))-\Big(\frac{\delta T^{(1)}_g(u)}{u}+\frac{\varepsilon_+g_1(u)}{2u}\Big)(2y-\delta(\bar y))\Big].
\end{eqnarray}
Consequently, the worm-gear L function is
\begin{eqnarray}\label{th:h1L-via-T}
h_{1L}^\perp(x,\vec b)&=&-x \int_{-1}^1 du \int_0^1 dy \delta (x-uy)y\Big(h_1(u)+\delta T_g^{(2)}(u)-\frac{\delta T_g^{(1)}(u)}{u}-\frac{\varepsilon_+g_1(u)}{2u}\Big)+O(\vec b^2).
\end{eqnarray}
Let us point that the expressions for worm-gear L and worm-gear T functions has very similar structure, compare eq.~(\ref{th:h1L-via-T}) to eq.~(\ref{th:g1T-via-T}). The main difference is the factor $y$ that appears in eq.~(\ref{th:g1T-via-T}). The integral over $\delta$-function could be evaluated with the result
\begin{eqnarray}
h_{1L}^\perp(x,\vec b)=-x^2 \int_x^1 \frac{du}{u^2}\(h_1(u)+\delta T_g^{(2)}(u)-\frac{\delta T_g^{(1)}(u)}{u}+\frac{\varepsilon_+g_1(u)}{2u}\)+O(\vec b^2),\qquad \text{for }x>0,
\\
h_{1L}^\perp(x,\vec b)=x^2 \int_{-1}^x \frac{du}{u^2}\(h_1(u)+\delta T_g^{(2)}(u)-\frac{\delta T_g^{(1)}(u)}{u}+\frac{\varepsilon_+g_1(u)}{2u}\)+O(\vec b^2),\qquad \text{for }x<0.
\end{eqnarray}
This expression can be compared to the first transverse momentum moment of TMD distribution derived in ref.~\cite{Kanazawa:2015ajw} (see eq.(49) in this reference). We  find that our expression agrees with the result of ref.~\cite{Kanazawa:2015ajw}.

\section{Mellin moments of worm-gear functions}
\label{sec:MMworm}

The final expressions for worm-gear function, as well as, all intermediate expressions are naturally expressed via Mellin convolutions. This fact  suggests a simple form for the Mellin moments of the worm-gear functions, which we present in this section.

First of all, let us point that functions $T^{(n)}$, $\Delta T^{(n)}$ and $\delta T^{(n)}$ defined in eq.~(\ref{def:Tn},~\ref{def:deltaTng}) obey  certain relations which simplify  in the algebra of Mellin moments. The Mellin moment of $T^{(n)}$ is
\begin{eqnarray}
\int_{-1}^1 dx x^k T^{(n)}(x)&=&\left\{\begin{array}{rl}
0,& k=0,\quad n \text{ odd},\\
0,& k\text{ odd},\\
T^{(n,k)}, &\text{otherwise},
\end{array}\right.
\end{eqnarray}
which  follows from the symmetry properties eq.~(\ref{sym_prop1}). The same relations hold for the function $\delta T_\epsilon$. For antisymmetric functions $\Delta T$ and $\delta T_g$ we have
\begin{eqnarray}
\int_{-1}^1 dx x^k \Delta T^{(n)}(x)&=&\left\{\begin{array}{rl}
0,& k=0,\quad n \text{ even},\\
0,& k\text{ even},\\
\Delta T^{(n,k)}, &\text{otherwise},
\end{array}\right.
\end{eqnarray}
which  follows from symmetry property eq.~(\ref{sym_prop2}). 

Evaluating the Mellin moments of the worm-gear functions in eq.~(\ref{th:gT_final}) and (\ref{th:hL_final}) we find simple expression
\begin{eqnarray}\label{mellin:g1T}
g^{(k)}_{1T}(\vec b)&=&\int_{-1}^1 dx x^k g_{1T}(x,\vec b)=\frac{1}{k+2}\left\{\begin{array}{ll}
g_1^{(k+1)}-\frac{1}{2}\Delta T^{(1,k)},& k\text{ odd},
\\
g_1^{(k+1)}+\Delta T^{(2,k+1)}+\frac{1}{2}T^{(1,k)},& k\text{ even},
\end{array}\right.
\end{eqnarray}
\begin{eqnarray}
h^{\perp(k)}_{1L}(\vec b)&=&\int_{-1}^1 dx x^k h_{1L}(x,\vec b)=\frac{-1}{k+3}\left\{\begin{array}{ll}
h_1^{(k+1)}-\delta T_g^{(1,k)},& k\text{ odd},
\\
h_1^{(k+1)}+\delta T_g^{(2,k+1)},& k\text{ even},
\end{array}\right.
\end{eqnarray}
where we omit the quark mass correction. The peculiar feature of this expressions is that functions with odd and even index $n$ enter different moments independently. 

Such relations can be important for lattice studies of TMD distributions, where only Mellin moments of distributions can be evaluated. For example, in ref.~\cite{Yoon:2017qzo} the lattice calculation of the first moment of $g_{1T}$ is performed. It has been found that 
\begin{eqnarray}\label{lattice1}
\frac{g_{1T}^{(0)}(\vec b\simeq 0.34)}{f_1^{(0)}(\vec b\simeq 0.34)}\Bigg|_{\text{\cite{Yoon:2017qzo}}}\approx 0.2.
\end{eqnarray}
The calculation has been done for the isovector combination of operators $q=u-d$. Here, the scales of TMD distributions are not clear since the translation rules between lattice scales and TMD evolution scales are not elaborated so far. Nonetheless, the evolution factors for both distributions are the same, and up to the first  order of approximation the scale dependence of eq.~(\ref{lattice1}) can be omitted. 
In ref.~\cite{Yoon:2017qzo} it is shown that $b$-dependence of the ratio in eq.~(\ref{lattice1}) is very weak. In particular, the value at $b\simeq 0.46$ practically coincides with eq.~(\ref{lattice1}). This suggests that the small-b expansion is a good approximation.

We can estimate the ratio in eq.~(\ref{lattice1}) from our calculation. Using eq.~(\ref{mellin:g1T}) we find
\begin{eqnarray}\label{lattice2}
\frac{g_{1T}^{(0)}(\vec b)}{f_1^{(0)}(\vec b)}=\frac{g_1^{(1)}+\Delta T^{(2,1)}}{2 f^{(0)}_1}+O(\alpha_s)+O(\vec b^2).
\end{eqnarray}
Assuming that the contribution of $\Delta T^{(2,1)}$ is small, i.e. in the Wandzura-Wilczek approximation, we find this ratio is $\sim 0.13$, which at this level of comparison is a good agreement.

\section{Conclusion}
\label{sec:results}

\begin{table}[t]
\begin{tabular}{l|c||c|c||c|c|c}
 		& 			& Leading 	& Twist of 	& Maximum 			&  		& Mix\\
Name 	& Function 	& matching 	& leading 	& known order 		& Ref. 	& with\\
		& 			& function 	& matching 	& of coef.function	& 		& gluon
\\\hline\hline
unpolarized & $f_1(x,\vec b)$ & $f_1$ &tw-2 & NNLO ($a_s^2$) & \cite{Gehrmann:2014yya,Echevarria:2016scs} & yes
\\\hline
Sivers & $f_{1T}^\perp(x,\vec b)$ & $T$ & tw-3 & LO ($a_s^0$) & \cite{Boer:2003cm,Ji:2006ub,Kang:2011mr}*$^\dagger$ eq.~(\ref{result:Sivers}) &yes
\\\hline\hline
helicity & $g_{1L}(x,\vec b)$ & $g_1$ & tw-2 & NLO ($a_s^1$) & \cite{Bacchetta:2013pqa,Gutierrez-Reyes:2017glx,Buffing:2017mqm} &yes
\\\hline
worm-gear T & $g_{1T}(x,\vec b)$ & $g_1$, $T$, $\Delta T$ & tw-2/3 & LO ($a_s^0$) &\cite{Kanazawa:2015ajw}* eq.~(\ref{result:wgT}) & yes
\\\hline\hline
transversity & $h_1(x,\vec b)$ & $h_1$& tw-2 & NNLO($a_s^2$)  & [NNLO] \cite{Gutierrez-Reyes:2018iod} & no
\\\hline
Boer-Mulders & $h_{1}^\perp(x,\vec b)$ & $\delta T_\epsilon$ & tw-3 & LO ($a_s^0$) & eq.~(\ref{result:Boer-Mulders}) & no
\\\hline
worm-gear L & $h_{1L}^\perp(x,\vec b)$ & $h_1$, $\delta T_g$ & tw-2/3 & LO ($a_s^0$) &\cite{Kanazawa:2015ajw}* eq.~(\ref{result:wgL}) & no
\\\hline
pretzelosity** & $h_{1T}^\perp(x,\vec b)$ &  -- & tw-4 & --  & -- & --
\end{tabular}
\\
~
\\
*~~The calculation is done in the momentum space. The result is given for the moments of distribution.
\\
$^\dagger$~~The calculation is done for the cross-section, with successive re-factorization into TMD distributions.
\\
**~~ The leading matching for the pretzelosity is currently unknown. We expect it to be of twist-4 level, \\see also discussion in \cite{Gutierrez-Reyes:2018iod}.
\caption{\label{tab:final-table} The summary of the quark TMD distributions and their leading matching at small-b.}
\end{table}

In this work we have evaluated the operator product expansion for the quark TMD operators up to linear in $\vec b$ terms. This order of expansion includes the majority of the polarized distributions. The summary of the matching relations is presented in  table~\ref{tab:final-table}. The main result of this study is the leading matching of Sivers, Boer-Mulders and worm gear function. We resume all the matching here for simplicity
\begin{eqnarray}\label{result:Sivers}
f_{1T}^\perp(x,\vec b)&=&\pm \pi T(-x,0,x)+O(\vec b^2),
\\\label{result:wgT}
g_{1T}(x,\vec b)&=&x\int_{-1}^{1} du \int_0^1 dy \delta(x-uy)\(g_1(u)+\Delta T^{(2)}(u)+\frac{T^{(1)}(u)-\Delta T^{(1)}(u)}{2u}\)+O(\vec b^2),
\\\label{result:Boer-Mulders}
h_1^\perp(x,\vec b)&=&\mp\pi \delta T_\epsilon(-x,0,x)+O(\vec b^2),
\\\label{result:wgL}
h_{1L}^\perp(x,\vec b)&=&-x \int_{-1}^1 du \int_0^1 dy \delta (x-uy)y\Big(h_1(u)+\delta T_g^{(2)}(u)-\frac{\delta T_g^{(1)}(u)}{u}\Big)+O(\vec b^2),
\end{eqnarray}
where the upper sign corresponds to the Drell-Yan operator, and lower sign corresponds to the SIDIS operator. The functions $g_1$ and $h_1$ are helicity and transversity PDFs. The functions $T$ are collinear distributions of twist-3. Their definition is given in (\ref{def:PDF_T}-\ref{def:PDF_deltaT}, \ref{def:Tn}-\ref{def:deltaTng}).

The expressions presented here are only the leading order perturbative QCD terms. The sub-leading terms include the power corrections in $\vec b^2$ and perturbative corrections. The perturbative corrections can be accumulated into the coefficient functions. For the distributions that match solely to twist-2 PDF these coefficient functions are already known at higher perturbative orders; next-to-next-to-leading order (NNLO) for unpolarized \cite{Gehrmann:2014yya,Echevarria:2015usa,Echevarria:2016scs} and transversity \cite{Gutierrez-Reyes:2018iod} distributions) and at NLO for helicity distribution \cite{Bacchetta:2013pqa,Gutierrez-Reyes:2017glx}. The polarized TMD distributions such as Sivers, Boer-Mulders, Collins and worm-gear functions, matches the twist-3 and twist-2 distributions. For these distributions, the coefficient functions are known only at LO and presented here altogether. 

We find that the results obtained by us agree with expressions that we have found in the literature (as far as we can trace necessary definitions of various components). However, there are several essential differences since all known expressions are given in the momentum space and they are presented in terms of certain integrals of TMD distribution over $p_T$. This fact complicates the comparison since such integrals are not well-defined within perturbation theory, and require some regularization procedure. In contrast, our calculation is done directly in the position space, and in this aspect, it represents a complete novelty. Another important distinctive fact of our work is that our calculation is based solely on the definition of TMD operators, whereas the majority of higher-twist calculations are based on the evaluation of particular cross-sections with successive re-interpretation in terms of TMDs. The only known example that we have found of a direct calculation is the one presented in ref.~\cite{Kanazawa:2015ajw}, where the leading matching for worm-gear functions is calculated. Finally, we consider all TMD distributions on the same foot and in the same framework which provides  a consistent relative normalization of all distributions improving their comprehension.

\subsection*{Acknowledgments}  
We greatly acknowledge V.Braun and A.Manashov for numerous discussions on details of higher twist calculus. We also thank Y. Koike for pointing out a mistake in the quark mass term that was in the initial version of the manuscript. I.S. is supported by the Spanish MECD grant FPA2016-75654-C2-2-P and the group UPARCOS. 

\appendix 

\section{Relation between TMD distributions in momentum and coordinate spaces}
\label{app:p<->b}

The momentum and coordinate representations are related by Fourier transformation (\ref{def:p<->b}),
\begin{eqnarray}
\Phi^{[\Gamma]}_{q\ot h}(x,p_T)=\int \frac{d^2 \vec b}{(2\pi)^2}e^{+i(\vec b\vec p_T)}\Phi^{[\Gamma]}_{q\ot h}(x,\vec b).
\end{eqnarray}
Performing the Fourier transformation of the parameterizations in eq.~(\ref{param:TMDv},~\ref{param:TMDa},\ref{param:TMDt}) and comparing it to the parameterizations in eq.~(\ref{def:gammaP_momentum},~\ref{def:gamma5_momentum},~\ref{def:sigma_momentum}) we find the relation between momentum and position space representations. They are conventionally presented using
\begin{eqnarray}
\widehat F^{(n)}(x,p_T)&=&\frac{M^{2n}}{n!}\int_0^\infty \frac{|\vec b|d|\vec b|}{2\pi}\(\frac{|\vec b|}{|\vec p_T|}\)^nJ_n(|\vec b||\vec p_T|)F(x,\vec b).
\end{eqnarray}
The inverse transformation is
\begin{eqnarray}
\widehat F^{(n)}(x,\vec b)=2\pi\frac{n!}{M^{2n}}\int_0^\infty |\vec p_T|d|\vec p_T|\(\frac{|\vec p_T|}{|\vec b|}\)^nJ_n(|\vec b||\vec p_T|)F(x,p_T).
\end{eqnarray}
Correspondingly, all TMDPDFs are split into three classes which transforms in the same manner,
\begin{eqnarray}
f_1=\widehat f^{(0)}_1,\qquad g_{1L}=\widehat g^{(0)}_{1L},\qquad h_1=\widehat h^{(0)}_1,
\end{eqnarray}
\begin{eqnarray}
f_{1T}^\perp=\widehat f_{1T}^{\perp(1)},\qquad g_{1T}=\widehat g^{(1)}_{1T},\qquad h_{1L}^\perp=\widehat h_{1L}^{\perp(1)},\qquad h_{1}^\perp=\widehat h_{1}^{\perp(1)},
\end{eqnarray}
\begin{eqnarray}
h_{1T}^\perp=\widehat h_{1T}^{\perp(2)}.
\end{eqnarray}

\section{Matrix element off-light cone}
\label{app:offLight}

The rules for working with matrix element off light-cone are discussed in details in \cite{Ball:1998ff,Ball:1998sk}. For completeness  we present here the intermediate steps which lead to equations (\ref{derO:v},~\ref{derO:a},~\ref{derO:t}). 

The initial step is the parameterization of matrix element of the operator off light-cone, in terms of four-dimensional vectors, $y^\mu$, $P^\mu$ and $S^\mu$, as well as, tensors $g^{\mu\nu}$ and $\epsilon^{\mu\nu\rho\sigma}$. 
Naturally, such a parameterization structurally repeats the parameterization of light-cone matrix element (\ref{param:vPDF},~\ref{param:aPDF},~\ref{param:tPDF}), with the replacement $p\to P$, $z\to y$ and $s_T\to S$,
\begin{eqnarray}\label{app:vPDF}
\langle P,S|O^{\gamma^\mu}(y)|P,S\rangle&=&2\int dx e^{2ix(yP)}\Big\{P^\mu A_1(x)+\frac{y^\mu}{(yP)}M^2 A_3(x)\Big\},
\\\label{app:aPDF}
\langle P,S|O^{\gamma^\mu\gamma^5}(y)|P,S\rangle&=&2\int dx e^{2ix(yP)}\Big\{P^\mu \frac{(yS)}{(yP)} M B_1(x)+S^\mu M B_2(x)+\frac{(yS)}{(yP)} M^3 \frac{y^\mu}{(yP)}B_3(x)\Big\},
\\\label{app:tPDF}
\langle P,S|O^{i\sigma^{\mu\nu}\gamma^5}(y)|P,S\rangle&=&2 \int dx e^{2ix(yP)}\Big\{(S^\mu P^\nu-P^\mu S^\nu) C_1(x)+\frac{(yS)}{(yP)^2} M^2(P^\mu y^\nu-y^\mu P^\nu)C_2(x)
\\\nn &&\qquad\qquad\qquad\qquad\qquad\qquad\qquad+(S^\mu y^\nu-y^\mu S^\nu)\frac{M^2}{(yP)} C_3(x)\Big\}.
\end{eqnarray}
The parameterization (\ref{app:vPDF},~\ref{app:tPDF}) is given in the space of physical vectors $(P^\mu,S^\mu,y^\mu)$ whereas the parameterization (\ref{param:vPDF},~\ref{param:aPDF},~\ref{param:tPDF}) is given in the space of light-cone vectors $(p^\mu,s_T^\mu,n^\mu)$. 
To connect these parameterizations we should relate the factorization frame to the physical frame. Assuming that $y^\mu \to z n^\mu$ in the limit $y^2\to0$, we obtain the following decomposition of $n^\mu$
\begin{eqnarray}\label{app:z->y}
z n^\mu=y^\mu-\frac{P^\mu}{M^2}\((yP)-\sqrt{(yP)^2-y^2 M^2}\).
\end{eqnarray}
Using this relation we decompose the vector $y^\mu$ over basis of $(p^\mu,s_T^\mu,n^\mu)$ and the small parameter $y^2$,
\begin{eqnarray}
y^\mu=z\Big[\frac{n^\mu}{2}\(1+\sqrt{1+\frac{y^2 M^2}{z^2(np)^2}}\)-\frac{p^\mu}{M^2}(np)\(1-\sqrt{1+\frac{y^2 M^2}{z^2(np)^2}}\)\Big]
\end{eqnarray}
The momentum and spin vectors are given by the definitions in eq.~(\ref{def:pmu}) and (\ref{def:Smu}),
\begin{eqnarray}
P^\mu&=&p^\mu+\frac{n^\mu}{2}\frac{M^2}{(np)},
\\
S^\mu&=&\frac{\lambda}{M}p^\mu-\frac{\lambda}{2}\frac{M}{(np)}n^\mu+s_T^\mu.
\end{eqnarray}
Therefore, the scalar products with the vector $y^\mu$ are
\begin{eqnarray}
(yP)&=&z(np)\sqrt{1+\frac{y^2M^2}{z^2(np)^2}},
\\
(yS)&=&\frac{\lambda}{M}z(np).
\end{eqnarray}
Using this translation dictionary we can compare the parameterizations (\ref{app:vPDF},~\ref{app:tPDF}) and (\ref{param:vPDF},~\ref{param:aPDF},~\ref{param:tPDF}) order-by-order in the parameter $y^2$. At the order $O(y^2)$ we obtain
\begin{eqnarray}\label{app:A=q}
&&  A_1=f_1,\qquad A_3=f_4-\frac{f_1}{2},
\\ &&
\label{app:B=q}
 B_1=g_1-g_T,\qquad B_2=g_T,\qquad B_3=g_3+g_T-\frac{g_1}{2},
\\ &&
\label{app:C=q}
 \nn C_1=h_1,\qquad C_2=h_L-h_3-\frac{h_1}{2},\qquad C_3=h_3-\frac{h_1}{2},
\end{eqnarray}
where we omit arguments of functions $(x)$ on both sides.

Obviously, the generalization off-light-cone is not unique. In particular, one can add terms power-suppressed in $y^2$ terms, to the definition in eq.~(\ref{app:z->y}). However, the reparameterization affects all intermediate steps of calculation and the difference should disappear in the final definite geometrical twist composition.  On top of this, such modifications are invisible at our level of accuracy.

\section{Equation of motion and functions $g_T$ and $h_L$}
\label{app:EOM}

The functions $g_T$ and $h_L$ are reducible, in the sense that they are compositions of \textit{geometrical twist}-2 functions ($g_1$ and $h_1$) and \textit{geometrical twist}-3 functions ($T$, $\Delta T$ and $\delta T_g$). The decomposition can be found with the help of the QCD equations of motion $\fnot D q=m q$. The convenient technique is described in \cite{Ball:1998ff} using  the example of distribution amplitudes. In the case of parton distributions the case is even simpler, since the total derivative contribution drops out,
$$\langle P,S|\partial_{\mu}\{...\}|P,S \rangle=0,$$
where dots represent any operator. In this appendix we present the derivation of functions $g_T$ and $h_L$ in details.

\subsection{Function $g_T$}
\label{app:gT}

As it is demonstrated in ref.~\cite{Ball:1998ff} the relation between $g_T$ and definite twist functions is found with the help of the following operator relation
\begin{eqnarray}\label{app:EOM1-op}
\bar q(y)\gamma_\mu\gamma^5[y,-y] q(-y)&=& \int_0^1 dt \frac{\partial}{\partial y^\mu} \bar q(ty)\fnot y \gamma^5[ty,-ty]q(-ty)
\\\nn&&-g\int_0^1 dt t \int_{-t}^t dv \frac{\epsilon_{\mu\nu\sigma\rho}y^\nu}{2}\bar q(ty)[ty,vy]F^{\sigma\rho}\fnot y[vy,-ty]q(-ty)
\\\nn&&-ig\int_0^1 dt\int_{-t}^t dv v \bar q(ty)[ty,vy]F_{\mu\nu}y^\nu\fnot y\gamma^5[vy,-ty]q(-ty)
\\\nn &&-i\epsilon_{\mu\nu\rho\sigma}\int_0^1 dt t y^\nu \partial^\rho\{\bar q(ty)\gamma^\sigma [ty,-ty]q(-ty)
\\\nn && +2m y^\nu \int_0^1 dt t \,\bar q(ty)\sigma_{\nu\mu}\gamma^5[ty,-ty] q(-ty)\},
\end{eqnarray}
where $m$ is the average mass of quark (if quark and anti-quark have different masses, it should be replaced by $m=(m_{\bar q}+m_q)/2$). This is an exact operator relation, and is the consequence of QCD equations of motion~\cite{Balitsky:1987bk}. Next, we evaluate the forward matrix element of equation (\ref{app:EOM1-op}) using the parameterizations (\ref{app:aPDF},~\ref{def:PDF_T},~\ref{def:PDF_DeltaT}). This operation transfers the variable $y^\mu$ into the elementary function, and the derivative over $y^\mu$ can be done. Next we take the limit $y^2\to 0$ as it is described in sec.~\ref{app:offLight}, and apply eq.~(\ref{app:B=q}). After that procedure we obtain the vector equation which contains the functions of different twists. Its $(2Ms_T^\mu)$-component is
\begin{eqnarray}&&\label{app:eom1}
\int du e^{2ix z p^+}g_T(x)=\int_0^1dt \int dx e^{2ixt zp^+}g_1(x)+i\varepsilon_+(zp^+)\int_0^1dt \int dx e^{2ixt zp^+} t \,h_1(x)
\\\nn &&\qquad\qquad+(zp^+)^2\int_0^1dt \int_{-t}^t dv v \Delta \tilde T(tz,vz,-tz)
-(zp^+)^2\int_0^1dt t \int_{-t}^t dv \tilde T(tz,vz,-tz),
\end{eqnarray}
where $\varepsilon_+=2m/M$. This equation relates collinear twist-3 function $g_T$ to the functions with geometrical twist 2 and 3.

To obtain the function $g_T$ explicitly, we perform the Fourier transformation for the equation (\ref{app:eom1}). It is convenient to write the result in the following form
\begin{eqnarray}\label{app:gT2}
g_T(x)&=&\int_0^1 dy \int_{-1}^1 du \delta(x-yu)\Big\{g_1(u) - \varepsilon_+\frac{h_1(u)}{2u}(1-\delta(\bar y))\Big\}
\\&&\nn
+\int_0^1 dy\int [dx]\Big\{\frac{1}{2}T(x_1,x_2,x_3)
\Big[\frac{\delta(x-x_3y)(1-\delta(\bar y))}{x_2x_3}+\frac{\delta(x+x_1y)(1-\delta(\bar y))}{x_1x_2}\Big]
\\&&\nn
-\frac{1}{2}\Delta T(x_1,x_2,x_3)
\Big[\frac{\delta(x-x_3y)(1-\delta(\bar y))}{x_2x_3}-\frac{\delta(x+x_1y)(1-\delta(\bar y))}{x_1x_2}\Big]
\\&&\nn
+\frac{\Delta T(x_1,x_2,x_3)}{x_2^2}
\Big[\delta(x-yx_3)-\delta(x+yx_1)\Big]\Big\}.
\end{eqnarray}
In this form it is simple to check the Burkhard-Cottingham sum rule
\begin{eqnarray}\label{app:g1_to_gT}
\int_{-1}^1 dx g_T(x)=\int_{-1}^1 dx g_1(x).
\end{eqnarray}
The equation (\ref{app:gT2}) has natural substructures in the form of $x_2$-moments introduced in eq.~(\ref{def:Tn},~\ref{def:DeltaTn}). Using the notation in eq.~(\ref{def:Tn},~\ref{def:DeltaTn}) we present the function $g_T$ as a Mellin convolution integral
\begin{eqnarray}\label{app:gT_final}
g_T(x)&=&\int_0^1 dy \int_{-1}^1 du \delta(x-yu)\Big[g_1(u)+\Delta T^{(2)}(u)+\frac{T^{(1)}(u)-\Delta T^{(1)}(u)-\varepsilon_+h_1(u)}{2u}(1-\delta(1-y))\Big].
\end{eqnarray}
Using this notation and the associativity of Mellin convolution it is simple to take the integral in eq.~(\ref{th:int_g1-gT}). It reads
\begin{eqnarray}\label{app:gT_convolution}
\int_{-1}^1 du \int_0^1 dy u(g_1(u)-g_T(u))\delta(x-yu)&=&
\int_{-1}^1 du \int_0^1 dy \delta(x-yu)\Big[ uy g_1(u)
\\\nn && \qquad-2\bar y u \Delta T^{(2)}(u)+y (T^{(1)}(u)-\Delta T^{(1)}(u)-\varepsilon_+h_1(u))\Big].
\end{eqnarray}

\subsection{Function $h_L$}
\label{app:hL}

The convenient form of the equation of motions for the derivation of function $h_L$ as given in ref.~\cite{Ball:1998ff}, 
\begin{eqnarray}
\label{app:EOM2-op}
\frac{\partial}{\partial y^\mu}\{\bar q(y)[y,-y](i\sigma^{\mu\nu}\gamma^5)y_\nu q(-y)\}&=&
ig\int_{-1}^1 dv\, v y^\nu y_\alpha \bar q(y)[y,v y]F_{\mu \nu}(v y)(i\sigma^{\mu\alpha}\gamma^5)[v y,-y]q(-y)
\\&& + y^\nu \partial_{\nu}\{\bar q(y)[y,-y]\gamma^5 q(-y)-2im\,\bar q(y)[y,-y]\fnot y\gamma^5 q(-y)\}.
\end{eqnarray}
Making the same steps as in the evaluation of the function $g_T$, i.e. considering matrix element with parameterizations as in eq.~(\ref{app:tPDF}) and (\ref{def:PDF_deltaT}), taking derivative and limit $y^2\to 0$, we obtain
\begin{eqnarray}\label{app:eom2}
\int dx e^{2ix zp^+}\[-ix zp^+ h_L(x)+(h_1(x)-h_L(x))\]=-(zp^+)^2\int_{-1}^1 dv\,v \delta \tilde T_g(z,v z,-z)
-\frac{i\varepsilon_+}{2}(zp^+)\int dx e^{2ixzp^+}g_1(x).
\end{eqnarray}
The Fourier transform of this equation leads to the differential equation
\begin{eqnarray}\label{app:diffeq_hL}
x\partial_x h_L(x)-h_L(x)+2h_1(x)&=&-4 p^+ \int \frac{dz}{2\pi} e^{-2ix z p^+} \int_{-1}^1 dv v (zp^+)^2\delta \tilde T_g(z,vz,-z)+\frac{\varepsilon_+}{2}\partial_x g_1(x)
\\\nn &=&\partial_x \delta T^{(1)}_g(x)-2\delta T^{(2)}_g(x)+\frac{\varepsilon_+}{2}\partial_x g_1(x).
\end{eqnarray}
The solution of this differential equation is
$$
h_L(x)=\int_0^1 dy \int_{-1}^1 du \delta(x-y u) y (2h_1(u)-\text{RHS}(u)),
$$
where RHS denotes the right-hand side of eq.~(\ref{app:diffeq_hL}). Performing an integration by parts we obtain
\begin{eqnarray}\label{app:hL_final}
h_L(x)&=&\int_0^1 dy \int_{-1}^1 du \delta(x-y u)\Big[2y(h_1(u)+\delta T^{(2)}_g(u))-\frac{\delta T^{(1)}_g(u)+\frac{\varepsilon_+}{2}g_1(u)}{u}(2y-\delta(1-y))\Big].
\end{eqnarray}
Clearly, it satisfies the Burkhard-Cottingham sum rule
\begin{eqnarray}\label{app:h1_to_hL}
\int_{-1}^1 dx h_L(x)=\int_{-1}^1 dx h_1(x).
\end{eqnarray}
It is intriguing to observe that the expression for $h_L$ (\ref{app:hL_final}) is structurally very similar to the expression for $g_T$ in eq.~(\ref{app:gT_final}).

\bibliography{TMD_ref}

\begin{thebibliography}{48}
\expandafter\ifx\csname natexlab\endcsname\relax\def\natexlab#1{#1}\fi
\expandafter\ifx\csname bibnamefont\endcsname\relax
  \def\bibnamefont#1{#1}\fi
\expandafter\ifx\csname bibfnamefont\endcsname\relax
  \def\bibfnamefont#1{#1}\fi
\expandafter\ifx\csname citenamefont\endcsname\relax
  \def\citenamefont#1{#1}\fi
\expandafter\ifx\csname url\endcsname\relax
  \def\url#1{\texttt{#1}}\fi
\expandafter\ifx\csname urlprefix\endcsname\relax\def\urlprefix{URL }\fi
\providecommand{\bibinfo}[2]{#2}
\providecommand{\eprint}[2][]{\url{#2}}

\bibitem[{\citenamefont{Collins}(2013)}]{Collins:2011zzd}
\bibinfo{author}{\bibfnamefont{J.}~\bibnamefont{Collins}},
  \emph{\bibinfo{title}{{Foundations of perturbative QCD}}}
  (\bibinfo{publisher}{Cambridge University Press}, \bibinfo{year}{2013}), ISBN
  \bibinfo{isbn}{9781107645257, 9781107645257, 9780521855334, 9781139097826},
  \urlprefix\url{http://www.cambridge.org/de/knowledge/isbn/item5756723}.

\bibitem[{\citenamefont{Echevarria et~al.}(2013)\citenamefont{Echevarria,
  Idilbi, and Scimemi}}]{Echevarria:2012js}
\bibinfo{author}{\bibfnamefont{M.~G.} \bibnamefont{Echevarria}},
  \bibinfo{author}{\bibfnamefont{A.}~\bibnamefont{Idilbi}}, \bibnamefont{and}
  \bibinfo{author}{\bibfnamefont{I.}~\bibnamefont{Scimemi}},
  \bibinfo{journal}{Phys. Lett.} \textbf{\bibinfo{volume}{B726}},
  \bibinfo{pages}{795} (\bibinfo{year}{2013}), \eprint{1211.1947}.

\bibitem[{\citenamefont{Echevarria et~al.}(2014)\citenamefont{Echevarria,
  Idilbi, and Scimemi}}]{Echevarria:2014rua}
\bibinfo{author}{\bibfnamefont{M.~G.} \bibnamefont{Echevarria}},
  \bibinfo{author}{\bibfnamefont{A.}~\bibnamefont{Idilbi}}, \bibnamefont{and}
  \bibinfo{author}{\bibfnamefont{I.}~\bibnamefont{Scimemi}},
  \bibinfo{journal}{Phys. Rev.} \textbf{\bibinfo{volume}{D90}},
  \bibinfo{pages}{014003} (\bibinfo{year}{2014}), \eprint{1402.0869}.

\bibitem[{\citenamefont{Vladimirov}(2018)}]{Vladimirov:2017ksc}
\bibinfo{author}{\bibfnamefont{A.}~\bibnamefont{Vladimirov}},
  \bibinfo{journal}{JHEP} \textbf{\bibinfo{volume}{04}}, \bibinfo{pages}{045}
  (\bibinfo{year}{2018}), \eprint{1707.07606}.

\bibitem[{\citenamefont{Scimemi and Vladimirov}(2018)}]{Scimemi:2017etj}
\bibinfo{author}{\bibfnamefont{I.}~\bibnamefont{Scimemi}} \bibnamefont{and}
  \bibinfo{author}{\bibfnamefont{A.}~\bibnamefont{Vladimirov}},
  \bibinfo{journal}{Eur. Phys. J.} \textbf{\bibinfo{volume}{C78}},
  \bibinfo{pages}{89} (\bibinfo{year}{2018}), \eprint{1706.01473}.

\bibitem[{\citenamefont{Echevarria
  et~al.}(2016{\natexlab{a}})\citenamefont{Echevarria, Scimemi, and
  Vladimirov}}]{Echevarria:2015byo}
\bibinfo{author}{\bibfnamefont{M.~G.} \bibnamefont{Echevarria}},
  \bibinfo{author}{\bibfnamefont{I.}~\bibnamefont{Scimemi}}, \bibnamefont{and}
  \bibinfo{author}{\bibfnamefont{A.}~\bibnamefont{Vladimirov}},
  \bibinfo{journal}{Phys. Rev.} \textbf{\bibinfo{volume}{D93}},
  \bibinfo{pages}{054004} (\bibinfo{year}{2016}{\natexlab{a}}),
  \eprint{1511.05590}.

\bibitem[{\citenamefont{Li and Zhu}(2017)}]{Li:2016ctv}
\bibinfo{author}{\bibfnamefont{Y.}~\bibnamefont{Li}} \bibnamefont{and}
  \bibinfo{author}{\bibfnamefont{H.~X.} \bibnamefont{Zhu}},
  \bibinfo{journal}{Phys. Rev. Lett.} \textbf{\bibinfo{volume}{118}},
  \bibinfo{pages}{022004} (\bibinfo{year}{2017}), \eprint{1604.01404}.

\bibitem[{\citenamefont{Vladimirov}(2017)}]{Vladimirov:2016dll}
\bibinfo{author}{\bibfnamefont{A.~A.} \bibnamefont{Vladimirov}},
  \bibinfo{journal}{Phys. Rev. Lett.} \textbf{\bibinfo{volume}{118}},
  \bibinfo{pages}{062001} (\bibinfo{year}{2017}), \eprint{1610.05791}.

\bibitem[{\citenamefont{Gutierrez-Reyes
  et~al.}(2017)\citenamefont{Gutierrez-Reyes, Scimemi, and
  Vladimirov}}]{Gutierrez-Reyes:2017glx}
\bibinfo{author}{\bibfnamefont{D.}~\bibnamefont{Gutierrez-Reyes}},
  \bibinfo{author}{\bibfnamefont{I.}~\bibnamefont{Scimemi}}, \bibnamefont{and}
  \bibinfo{author}{\bibfnamefont{A.~A.} \bibnamefont{Vladimirov}},
  \bibinfo{journal}{Phys. Lett.} \textbf{\bibinfo{volume}{B769}},
  \bibinfo{pages}{84} (\bibinfo{year}{2017}), \eprint{1702.06558}.

\bibitem[{\citenamefont{Echevarria
  et~al.}(2016{\natexlab{b}})\citenamefont{Echevarria, Scimemi, and
  Vladimirov}}]{Echevarria:2015usa}
\bibinfo{author}{\bibfnamefont{M.~G.} \bibnamefont{Echevarria}},
  \bibinfo{author}{\bibfnamefont{I.}~\bibnamefont{Scimemi}}, \bibnamefont{and}
  \bibinfo{author}{\bibfnamefont{A.}~\bibnamefont{Vladimirov}},
  \bibinfo{journal}{Phys. Rev.} \textbf{\bibinfo{volume}{D93}},
  \bibinfo{pages}{011502} (\bibinfo{year}{2016}{\natexlab{b}}),
  \bibinfo{note}{[Erratum: Phys. Rev.D94,no.9,099904(2016)]},
  \eprint{1509.06392}.

\bibitem[{\citenamefont{Echevarria
  et~al.}(2016{\natexlab{c}})\citenamefont{Echevarria, Scimemi, and
  Vladimirov}}]{Echevarria:2016scs}
\bibinfo{author}{\bibfnamefont{M.~G.} \bibnamefont{Echevarria}},
  \bibinfo{author}{\bibfnamefont{I.}~\bibnamefont{Scimemi}}, \bibnamefont{and}
  \bibinfo{author}{\bibfnamefont{A.}~\bibnamefont{Vladimirov}},
  \bibinfo{journal}{JHEP} \textbf{\bibinfo{volume}{09}}, \bibinfo{pages}{004}
  (\bibinfo{year}{2016}{\natexlab{c}}), \eprint{1604.07869}.

\bibitem[{\citenamefont{Gutierrez-Reyes
  et~al.}(2018)\citenamefont{Gutierrez-Reyes, Scimemi, and
  Vladimirov}}]{Gutierrez-Reyes:2018iod}
\bibinfo{author}{\bibfnamefont{D.}~\bibnamefont{Gutierrez-Reyes}},
  \bibinfo{author}{\bibfnamefont{I.}~\bibnamefont{Scimemi}}, \bibnamefont{and}
  \bibinfo{author}{\bibfnamefont{A.}~\bibnamefont{Vladimirov}},
  \bibinfo{journal}{JHEP} \textbf{\bibinfo{volume}{07}}, \bibinfo{pages}{172}
  (\bibinfo{year}{2018}), \eprint{1805.07243}.

\bibitem[{\citenamefont{Bacchetta et~al.}(2008)\citenamefont{Bacchetta, Boer,
  Diehl, and Mulders}}]{Bacchetta:2008xw}
\bibinfo{author}{\bibfnamefont{A.}~\bibnamefont{Bacchetta}},
  \bibinfo{author}{\bibfnamefont{D.}~\bibnamefont{Boer}},
  \bibinfo{author}{\bibfnamefont{M.}~\bibnamefont{Diehl}}, \bibnamefont{and}
  \bibinfo{author}{\bibfnamefont{P.~J.} \bibnamefont{Mulders}},
  \bibinfo{journal}{JHEP} \textbf{\bibinfo{volume}{08}}, \bibinfo{pages}{023}
  (\bibinfo{year}{2008}), \eprint{0803.0227}.

\bibitem[{\citenamefont{Meissner et~al.}(2009)\citenamefont{Meissner, Metz, and
  Schlegel}}]{Meissner:2009ww}
\bibinfo{author}{\bibfnamefont{S.}~\bibnamefont{Meissner}},
  \bibinfo{author}{\bibfnamefont{A.}~\bibnamefont{Metz}}, \bibnamefont{and}
  \bibinfo{author}{\bibfnamefont{M.}~\bibnamefont{Schlegel}},
  \bibinfo{journal}{JHEP} \textbf{\bibinfo{volume}{08}}, \bibinfo{pages}{056}
  (\bibinfo{year}{2009}), \eprint{0906.5323}.

\bibitem[{\citenamefont{Lorce et~al.}(2011)\citenamefont{Lorce, Pasquini, and
  Vanderhaeghen}}]{Lorce:2011dv}
\bibinfo{author}{\bibfnamefont{C.}~\bibnamefont{Lorce}},
  \bibinfo{author}{\bibfnamefont{B.}~\bibnamefont{Pasquini}}, \bibnamefont{and}
  \bibinfo{author}{\bibfnamefont{M.}~\bibnamefont{Vanderhaeghen}},
  \bibinfo{journal}{JHEP} \textbf{\bibinfo{volume}{05}}, \bibinfo{pages}{041}
  (\bibinfo{year}{2011}), \eprint{1102.4704}.

\bibitem[{\citenamefont{Ji et~al.}(2006)\citenamefont{Ji, Qiu, Vogelsang, and
  Yuan}}]{Ji:2006ub}
\bibinfo{author}{\bibfnamefont{X.}~\bibnamefont{Ji}},
  \bibinfo{author}{\bibfnamefont{J.-W.} \bibnamefont{Qiu}},
  \bibinfo{author}{\bibfnamefont{W.}~\bibnamefont{Vogelsang}},
  \bibnamefont{and} \bibinfo{author}{\bibfnamefont{F.}~\bibnamefont{Yuan}},
  \bibinfo{journal}{Phys. Rev. Lett.} \textbf{\bibinfo{volume}{97}},
  \bibinfo{pages}{082002} (\bibinfo{year}{2006}), \eprint{hep-ph/0602239}.

\bibitem[{\citenamefont{Kang et~al.}(2011{\natexlab{a}})\citenamefont{Kang,
  Xiao, and Yuan}}]{Kang:2011mr}
\bibinfo{author}{\bibfnamefont{Z.-B.} \bibnamefont{Kang}},
  \bibinfo{author}{\bibfnamefont{B.-W.} \bibnamefont{Xiao}}, \bibnamefont{and}
  \bibinfo{author}{\bibfnamefont{F.}~\bibnamefont{Yuan}},
  \bibinfo{journal}{Phys. Rev. Lett.} \textbf{\bibinfo{volume}{107}},
  \bibinfo{pages}{152002} (\bibinfo{year}{2011}{\natexlab{a}}),
  \eprint{1106.0266}.

\bibitem[{\citenamefont{Gehrmann et~al.}(2014)\citenamefont{Gehrmann, Luebbert,
  and Yang}}]{Gehrmann:2014yya}
\bibinfo{author}{\bibfnamefont{T.}~\bibnamefont{Gehrmann}},
  \bibinfo{author}{\bibfnamefont{T.}~\bibnamefont{Luebbert}}, \bibnamefont{and}
  \bibinfo{author}{\bibfnamefont{L.~L.} \bibnamefont{Yang}},
  \bibinfo{journal}{JHEP} \textbf{\bibinfo{volume}{06}}, \bibinfo{pages}{155}
  (\bibinfo{year}{2014}), \eprint{1403.6451}.

\bibitem[{\citenamefont{Collins}(2002)}]{Collins:2002kn}
\bibinfo{author}{\bibfnamefont{J.~C.} \bibnamefont{Collins}},
  \bibinfo{journal}{Phys. Lett.} \textbf{\bibinfo{volume}{B536}},
  \bibinfo{pages}{43} (\bibinfo{year}{2002}), \eprint{hep-ph/0204004}.

\bibitem[{\citenamefont{Efremov and Teryaev}(1984)}]{Efremov:1983eb}
\bibinfo{author}{\bibfnamefont{A.~V.} \bibnamefont{Efremov}} \bibnamefont{and}
  \bibinfo{author}{\bibfnamefont{O.~V.} \bibnamefont{Teryaev}},
  \bibinfo{journal}{Sov. J. Nucl. Phys.} \textbf{\bibinfo{volume}{39}},
  \bibinfo{pages}{962} (\bibinfo{year}{1984}), \bibinfo{note}{[Yad.
  Fiz.39,1517(1984)]}.

\bibitem[{\citenamefont{Efremov and Teryaev}(1985)}]{Efremov:1984ip}
\bibinfo{author}{\bibfnamefont{A.~V.} \bibnamefont{Efremov}} \bibnamefont{and}
  \bibinfo{author}{\bibfnamefont{O.~V.} \bibnamefont{Teryaev}},
  \bibinfo{journal}{Phys. Lett.} \textbf{\bibinfo{volume}{150B}},
  \bibinfo{pages}{383} (\bibinfo{year}{1985}).

\bibitem[{\citenamefont{Qiu and Sterman}(1991)}]{Qiu:1991pp}
\bibinfo{author}{\bibfnamefont{J.-w.} \bibnamefont{Qiu}} \bibnamefont{and}
  \bibinfo{author}{\bibfnamefont{G.~F.} \bibnamefont{Sterman}},
  \bibinfo{journal}{Phys. Rev. Lett.} \textbf{\bibinfo{volume}{67}},
  \bibinfo{pages}{2264} (\bibinfo{year}{1991}).

\bibitem[{\citenamefont{Qiu and Sterman}(1992)}]{Qiu:1991wg}
\bibinfo{author}{\bibfnamefont{J.-w.} \bibnamefont{Qiu}} \bibnamefont{and}
  \bibinfo{author}{\bibfnamefont{G.~F.} \bibnamefont{Sterman}},
  \bibinfo{journal}{Nucl. Phys.} \textbf{\bibinfo{volume}{B378}},
  \bibinfo{pages}{52} (\bibinfo{year}{1992}).

\bibitem[{\citenamefont{Tangerman and Mulders}(1995)}]{Tangerman:1994eh}
\bibinfo{author}{\bibfnamefont{R.~D.} \bibnamefont{Tangerman}}
  \bibnamefont{and} \bibinfo{author}{\bibfnamefont{P.~J.}
  \bibnamefont{Mulders}}, \bibinfo{journal}{Phys. Rev.}
  \textbf{\bibinfo{volume}{D51}}, \bibinfo{pages}{3357} (\bibinfo{year}{1995}),
  \eprint{hep-ph/9403227}.

\bibitem[{\citenamefont{Echevarria et~al.}(2012)\citenamefont{Echevarria,
  Idilbi, and Scimemi}}]{GarciaEchevarria:2011rb}
\bibinfo{author}{\bibfnamefont{M.~G.} \bibnamefont{Echevarria}},
  \bibinfo{author}{\bibfnamefont{A.}~\bibnamefont{Idilbi}}, \bibnamefont{and}
  \bibinfo{author}{\bibfnamefont{I.}~\bibnamefont{Scimemi}},
  \bibinfo{journal}{JHEP} \textbf{\bibinfo{volume}{07}}, \bibinfo{pages}{002}
  (\bibinfo{year}{2012}), \eprint{1111.4996}.

\bibitem[{\citenamefont{Mulders and Tangerman}(1996)}]{Mulders:1995dh}
\bibinfo{author}{\bibfnamefont{P.~J.} \bibnamefont{Mulders}} \bibnamefont{and}
  \bibinfo{author}{\bibfnamefont{R.~D.} \bibnamefont{Tangerman}},
  \bibinfo{journal}{Nucl. Phys.} \textbf{\bibinfo{volume}{B461}},
  \bibinfo{pages}{197} (\bibinfo{year}{1996}), \bibinfo{note}{[Erratum: Nucl.
  Phys.B484,538(1997)]}, \eprint{hep-ph/9510301}.

\bibitem[{\citenamefont{Bacchetta et~al.}(2007)\citenamefont{Bacchetta, Diehl,
  Goeke, Metz, Mulders, and Schlegel}}]{Bacchetta:2006tn}
\bibinfo{author}{\bibfnamefont{A.}~\bibnamefont{Bacchetta}},
  \bibinfo{author}{\bibfnamefont{M.}~\bibnamefont{Diehl}},
  \bibinfo{author}{\bibfnamefont{K.}~\bibnamefont{Goeke}},
  \bibinfo{author}{\bibfnamefont{A.}~\bibnamefont{Metz}},
  \bibinfo{author}{\bibfnamefont{P.~J.} \bibnamefont{Mulders}},
  \bibnamefont{and} \bibinfo{author}{\bibfnamefont{M.}~\bibnamefont{Schlegel}},
  \bibinfo{journal}{JHEP} \textbf{\bibinfo{volume}{02}}, \bibinfo{pages}{093}
  (\bibinfo{year}{2007}), \eprint{hep-ph/0611265}.

\bibitem[{\citenamefont{Goeke et~al.}(2005)\citenamefont{Goeke, Metz, and
  Schlegel}}]{Goeke:2005hb}
\bibinfo{author}{\bibfnamefont{K.}~\bibnamefont{Goeke}},
  \bibinfo{author}{\bibfnamefont{A.}~\bibnamefont{Metz}}, \bibnamefont{and}
  \bibinfo{author}{\bibfnamefont{M.}~\bibnamefont{Schlegel}},
  \bibinfo{journal}{Phys. Lett.} \textbf{\bibinfo{volume}{B618}},
  \bibinfo{pages}{90} (\bibinfo{year}{2005}), \eprint{hep-ph/0504130}.

\bibitem[{\citenamefont{Boer et~al.}(2011)\citenamefont{Boer, Gamberg, Musch,
  and Prokudin}}]{Boer:2011xd}
\bibinfo{author}{\bibfnamefont{D.}~\bibnamefont{Boer}},
  \bibinfo{author}{\bibfnamefont{L.}~\bibnamefont{Gamberg}},
  \bibinfo{author}{\bibfnamefont{B.}~\bibnamefont{Musch}}, \bibnamefont{and}
  \bibinfo{author}{\bibfnamefont{A.}~\bibnamefont{Prokudin}},
  \bibinfo{journal}{JHEP} \textbf{\bibinfo{volume}{10}}, \bibinfo{pages}{021}
  (\bibinfo{year}{2011}), \eprint{1107.5294}.

\bibitem[{\citenamefont{Belitsky et~al.}(2003)\citenamefont{Belitsky, Ji, and
  Yuan}}]{Belitsky:2002sm}
\bibinfo{author}{\bibfnamefont{A.~V.} \bibnamefont{Belitsky}},
  \bibinfo{author}{\bibfnamefont{X.}~\bibnamefont{Ji}}, \bibnamefont{and}
  \bibinfo{author}{\bibfnamefont{F.}~\bibnamefont{Yuan}},
  \bibinfo{journal}{Nucl. Phys.} \textbf{\bibinfo{volume}{B656}},
  \bibinfo{pages}{165} (\bibinfo{year}{2003}), \eprint{hep-ph/0208038}.

\bibitem[{\citenamefont{Idilbi and
  Scimemi}(2011{\natexlab{a}})}]{Idilbi:2010im}
\bibinfo{author}{\bibfnamefont{A.}~\bibnamefont{Idilbi}} \bibnamefont{and}
  \bibinfo{author}{\bibfnamefont{I.}~\bibnamefont{Scimemi}},
  \bibinfo{journal}{Phys. Lett.} \textbf{\bibinfo{volume}{B695}},
  \bibinfo{pages}{463} (\bibinfo{year}{2011}{\natexlab{a}}),
  \eprint{1009.2776}.

\bibitem[{\citenamefont{Idilbi and
  Scimemi}(2011{\natexlab{b}})}]{Idilbi:2010tc}
\bibinfo{author}{\bibfnamefont{A.}~\bibnamefont{Idilbi}} \bibnamefont{and}
  \bibinfo{author}{\bibfnamefont{I.}~\bibnamefont{Scimemi}},
  \bibinfo{journal}{AIP Conf. Proc.} \textbf{\bibinfo{volume}{1343}},
  \bibinfo{pages}{320} (\bibinfo{year}{2011}{\natexlab{b}}),
  \eprint{1012.4419}.

\bibitem[{\citenamefont{Garcia-Echevarria
  et~al.}(2011)\citenamefont{Garcia-Echevarria, Idilbi, and
  Scimemi}}]{GarciaEchevarria:2011md}
\bibinfo{author}{\bibfnamefont{M.}~\bibnamefont{Garcia-Echevarria}},
  \bibinfo{author}{\bibfnamefont{A.}~\bibnamefont{Idilbi}}, \bibnamefont{and}
  \bibinfo{author}{\bibfnamefont{I.}~\bibnamefont{Scimemi}},
  \bibinfo{journal}{Phys. Rev.} \textbf{\bibinfo{volume}{D84}},
  \bibinfo{pages}{011502} (\bibinfo{year}{2011}), \eprint{1104.0686}.

\bibitem[{\citenamefont{Belitsky and Mueller}(2000)}]{Belitsky:2000vx}
\bibinfo{author}{\bibfnamefont{A.~V.} \bibnamefont{Belitsky}} \bibnamefont{and}
  \bibinfo{author}{\bibfnamefont{D.}~\bibnamefont{Mueller}},
  \bibinfo{journal}{Nucl. Phys.} \textbf{\bibinfo{volume}{B589}},
  \bibinfo{pages}{611} (\bibinfo{year}{2000}), \eprint{hep-ph/0007031}.

\bibitem[{\citenamefont{Balitsky and Braun}(1989)}]{Balitsky:1987bk}
\bibinfo{author}{\bibfnamefont{I.~I.} \bibnamefont{Balitsky}} \bibnamefont{and}
  \bibinfo{author}{\bibfnamefont{V.~M.} \bibnamefont{Braun}},
  \bibinfo{journal}{Nucl. Phys.} \textbf{\bibinfo{volume}{B311}},
  \bibinfo{pages}{541} (\bibinfo{year}{1989}).

\bibitem[{\citenamefont{Ball et~al.}(1998)\citenamefont{Ball, Braun, Koike, and
  Tanaka}}]{Ball:1998sk}
\bibinfo{author}{\bibfnamefont{P.}~\bibnamefont{Ball}},
  \bibinfo{author}{\bibfnamefont{V.~M.} \bibnamefont{Braun}},
  \bibinfo{author}{\bibfnamefont{Y.}~\bibnamefont{Koike}}, \bibnamefont{and}
  \bibinfo{author}{\bibfnamefont{K.}~\bibnamefont{Tanaka}},
  \bibinfo{journal}{Nucl. Phys.} \textbf{\bibinfo{volume}{B529}},
  \bibinfo{pages}{323} (\bibinfo{year}{1998}), \eprint{hep-ph/9802299}.

\bibitem[{\citenamefont{Jaffe and Ji}(1992)}]{Jaffe:1991ra}
\bibinfo{author}{\bibfnamefont{R.~L.} \bibnamefont{Jaffe}} \bibnamefont{and}
  \bibinfo{author}{\bibfnamefont{X.-D.} \bibnamefont{Ji}},
  \bibinfo{journal}{Nucl. Phys.} \textbf{\bibinfo{volume}{B375}},
  \bibinfo{pages}{527} (\bibinfo{year}{1992}).

\bibitem[{\citenamefont{Braun et~al.}(2009)\citenamefont{Braun, Manashov, and
  Pirnay}}]{Braun:2009mi}
\bibinfo{author}{\bibfnamefont{V.~M.} \bibnamefont{Braun}},
  \bibinfo{author}{\bibfnamefont{A.~N.} \bibnamefont{Manashov}},
  \bibnamefont{and} \bibinfo{author}{\bibfnamefont{B.}~\bibnamefont{Pirnay}},
  \bibinfo{journal}{Phys. Rev.} \textbf{\bibinfo{volume}{D80}},
  \bibinfo{pages}{114002} (\bibinfo{year}{2009}), \bibinfo{note}{[Erratum:
  Phys. Rev.D86,119902(2012)]}, \eprint{0909.3410}.

\bibitem[{\citenamefont{Jaffe}(1983)}]{Jaffe:1983hp}
\bibinfo{author}{\bibfnamefont{R.~L.} \bibnamefont{Jaffe}},
  \bibinfo{journal}{Nucl. Phys.} \textbf{\bibinfo{volume}{B229}},
  \bibinfo{pages}{205} (\bibinfo{year}{1983}).

\bibitem[{\citenamefont{Burkhardt and Cottingham}(1970)}]{Burkhardt:1970ti}
\bibinfo{author}{\bibfnamefont{H.}~\bibnamefont{Burkhardt}} \bibnamefont{and}
  \bibinfo{author}{\bibfnamefont{W.~N.} \bibnamefont{Cottingham}},
  \bibinfo{journal}{Annals Phys.} \textbf{\bibinfo{volume}{56}},
  \bibinfo{pages}{453} (\bibinfo{year}{1970}).

\bibitem[{\citenamefont{Ball and Braun}(1999)}]{Ball:1998ff}
\bibinfo{author}{\bibfnamefont{P.}~\bibnamefont{Ball}} \bibnamefont{and}
  \bibinfo{author}{\bibfnamefont{V.~M.} \bibnamefont{Braun}},
  \bibinfo{journal}{Nucl. Phys.} \textbf{\bibinfo{volume}{B543}},
  \bibinfo{pages}{201} (\bibinfo{year}{1999}), \eprint{hep-ph/9810475}.

\bibitem[{\citenamefont{Kang and Qiu}(2009)}]{Kang:2008ey}
\bibinfo{author}{\bibfnamefont{Z.-B.} \bibnamefont{Kang}} \bibnamefont{and}
  \bibinfo{author}{\bibfnamefont{J.-W.} \bibnamefont{Qiu}},
  \bibinfo{journal}{Phys. Rev.} \textbf{\bibinfo{volume}{D79}},
  \bibinfo{pages}{016003} (\bibinfo{year}{2009}), \eprint{0811.3101}.

\bibitem[{\citenamefont{Boer et~al.}(2003)\citenamefont{Boer, Mulders, and
  Pijlman}}]{Boer:2003cm}
\bibinfo{author}{\bibfnamefont{D.}~\bibnamefont{Boer}},
  \bibinfo{author}{\bibfnamefont{P.~J.} \bibnamefont{Mulders}},
  \bibnamefont{and} \bibinfo{author}{\bibfnamefont{F.}~\bibnamefont{Pijlman}},
  \bibinfo{journal}{Nucl. Phys.} \textbf{\bibinfo{volume}{B667}},
  \bibinfo{pages}{201} (\bibinfo{year}{2003}), \eprint{hep-ph/0303034}.

\bibitem[{\citenamefont{Kang et~al.}(2011{\natexlab{b}})\citenamefont{Kang,
  Qiu, Vogelsang, and Yuan}}]{Kang:2011hk}
\bibinfo{author}{\bibfnamefont{Z.-B.} \bibnamefont{Kang}},
  \bibinfo{author}{\bibfnamefont{J.-W.} \bibnamefont{Qiu}},
  \bibinfo{author}{\bibfnamefont{W.}~\bibnamefont{Vogelsang}},
  \bibnamefont{and} \bibinfo{author}{\bibfnamefont{F.}~\bibnamefont{Yuan}},
  \bibinfo{journal}{Phys. Rev.} \textbf{\bibinfo{volume}{D83}},
  \bibinfo{pages}{094001} (\bibinfo{year}{2011}{\natexlab{b}}),
  \eprint{1103.1591}.

\bibitem[{\citenamefont{Kanazawa et~al.}(2016)\citenamefont{Kanazawa, Koike,
  Metz, Pitonyak, and Schlegel}}]{Kanazawa:2015ajw}
\bibinfo{author}{\bibfnamefont{K.}~\bibnamefont{Kanazawa}},
  \bibinfo{author}{\bibfnamefont{Y.}~\bibnamefont{Koike}},
  \bibinfo{author}{\bibfnamefont{A.}~\bibnamefont{Metz}},
  \bibinfo{author}{\bibfnamefont{D.}~\bibnamefont{Pitonyak}}, \bibnamefont{and}
  \bibinfo{author}{\bibfnamefont{M.}~\bibnamefont{Schlegel}},
  \bibinfo{journal}{Phys. Rev.} \textbf{\bibinfo{volume}{D93}},
  \bibinfo{pages}{054024} (\bibinfo{year}{2016}), \eprint{1512.07233}.

\bibitem[{\citenamefont{Yoon et~al.}(2017)\citenamefont{Yoon, Engelhardt,
  Gupta, Bhattacharya, Green, Musch, Negele, Pochinsky, Schäfer, and
  Syritsyn}}]{Yoon:2017qzo}
\bibinfo{author}{\bibfnamefont{B.}~\bibnamefont{Yoon}},
  \bibinfo{author}{\bibfnamefont{M.}~\bibnamefont{Engelhardt}},
  \bibinfo{author}{\bibfnamefont{R.}~\bibnamefont{Gupta}},
  \bibinfo{author}{\bibfnamefont{T.}~\bibnamefont{Bhattacharya}},
  \bibinfo{author}{\bibfnamefont{J.~R.} \bibnamefont{Green}},
  \bibinfo{author}{\bibfnamefont{B.~U.} \bibnamefont{Musch}},
  \bibinfo{author}{\bibfnamefont{J.~W.} \bibnamefont{Negele}},
  \bibinfo{author}{\bibfnamefont{A.~V.} \bibnamefont{Pochinsky}},
  \bibinfo{author}{\bibfnamefont{A.}~\bibnamefont{Schäfer}}, \bibnamefont{and}
  \bibinfo{author}{\bibfnamefont{S.~N.} \bibnamefont{Syritsyn}},
  \bibinfo{journal}{Phys. Rev.} \textbf{\bibinfo{volume}{D96}},
  \bibinfo{pages}{094508} (\bibinfo{year}{2017}), \eprint{1706.03406}.

\bibitem[{\citenamefont{Bacchetta and Prokudin}(2013)}]{Bacchetta:2013pqa}
\bibinfo{author}{\bibfnamefont{A.}~\bibnamefont{Bacchetta}} \bibnamefont{and}
  \bibinfo{author}{\bibfnamefont{A.}~\bibnamefont{Prokudin}},
  \bibinfo{journal}{Nucl. Phys.} \textbf{\bibinfo{volume}{B875}},
  \bibinfo{pages}{536} (\bibinfo{year}{2013}), \eprint{1303.2129}.

\bibitem[{\citenamefont{Buffing et~al.}(2018)\citenamefont{Buffing, Diehl, and
  Kasemets}}]{Buffing:2017mqm}
\bibinfo{author}{\bibfnamefont{M.~G.~A.} \bibnamefont{Buffing}},
  \bibinfo{author}{\bibfnamefont{M.}~\bibnamefont{Diehl}}, \bibnamefont{and}
  \bibinfo{author}{\bibfnamefont{T.}~\bibnamefont{Kasemets}},
  \bibinfo{journal}{JHEP} \textbf{\bibinfo{volume}{01}}, \bibinfo{pages}{044}
  (\bibinfo{year}{2018}), \eprint{1708.03528}.

\end{thebibliography}
\end{document}